\documentclass[a4paper,fleqn]{cas-sc}

\usepackage[authoryear,longnamesfirst]{natbib}
\usepackage{multirow}
\usepackage{tabularx}
\usepackage{tikz}
\usepackage{float}
\usepackage[section]{placeins}
\usepackage{booktabs}
\usepackage{graphicx}
\usepackage{url}
\usepackage{pdflscape}
\usepackage{longtable}
\usepackage{array}
\usepackage[table]{xcolor}
\usepackage{ragged2e}
\newcolumntype{L}[1]{>{\RaggedRight\arraybackslash}p{#1}}

\usetikzlibrary{shapes.geometric, arrows.meta, positioning, fit}

\begin{document}
\let\WriteBookmarks\relax

\shorttitle{HugSelect: Explainable foundation-model selection}
\shortauthors{Joonbakhsh et~al.}

\title[mode = title]{HugSelect: An Explainable Multi-Criteria Decision-Support Framework for foundation-model selection}


\author[1]{Alireza Joonbakhsh}[orcid=0000-0001-5044-5196]
\cormark[1]
\ead{alireza.joonbakhsh@hafez.shirazu.ac.ir}

\author[2]{Arda Canser Adalı}
\ead{a.c.adali@students.uu.nl}

\author[2]{Slinger Jansen}[orcid=0000-0003-3752-2868]
\ead{slinger.jansen@uu.nl}

\author[1]{Farshad Khunjush}[orcid=0000-0002-3339-6051]
\ead{khunjush@shirazu.ac.ir}

\author[3]{Siamak Farshidi}[orcid=0000-0001-6139-921X]
\cormark[1]
\ead{siamak.farshidi@wur.nl}

\cortext[1]{Corresponding authors: Alireza Joonbakhsh and Siamak Farshidi.}


\affiliation[1]{
    organization={Shiraz University},
    city={Shiraz},
    country={Iran}
}

\affiliation[2]{
    organization={Department of Information and Computing Sciences, Utrecht University},
    city={Utrecht},
    country={The Netherlands}
}

\affiliation[3]{
    organization={Information Technology Group, Wageningen University \& Research},
    city={Wageningen},
    country={The Netherlands}
}

\begin{abstract}
Foundation models are increasingly reused as software components, making model selection a critical software-engineering decision. Current model hubs primarily support discovery through popularity metrics, often neglecting functional capabilities, operational constraints, and community-perceived quality. We argue that foundation-model selection should be treated as an explicit, auditable software-component selection task rather than as keyword search, popularity ranking, or opaque conversational advice.

This paper proposes \textit{HugSelect}, an explainable decision-support framework for foundation-model selection. HugSelect builds a knowledge base of 71,274 models by combining repository metadata, extracted functional capabilities, and perceived quality attributes derived from community discussions into a unified pipeline. It ranks candidate models using a weighted additive model that exposes criterion-level score decompositions.

We evaluated HugSelect through pipeline validation, comparative case studies against four commercial LLM-based recommendation systems (44 scenarios), fine-grained ablation, and an exploratory user study ($n=10$). Extraction pipelines achieved an F1-score of 0.801 for functional features and 0.84 accuracy for quality-attribute mapping. HugSelect achieved a model-level Coverage@10 of 0.61 and family-level Coverage@10 of 0.91, showing recommendation quality comparable to the evaluated commercial systems without significant overall differences in ranking quality, while providing stable, traceable, and inspectable reasoning. Ablation confirmed functional features as the main driver of retrieval accuracy, and preliminary user feedback suggests the framework is useful and intuitive.

\end{abstract}


\maketitle

\section{Introduction}
\label{sec:introduction}

\subsection*{Motivation and problem context}

foundation-models have become reusable building blocks for AI-enabled software systems. They are no longer only research artifacts: practitioners increasingly integrate them into applications that perform text generation, question answering, visual understanding, code assistance, recommendation, and decision support~\cite{bommasani2021opportunities,cazzaniga2024gen}. In this setting, choosing a foundation-model resembles selecting a complex third-party software component. The decision influences whether the resulting system satisfies functional requirements, but it also affects non-functional properties such as performance efficiency, reliability, maintainability, security, deployment effort, licensing constraints, and operational cost.

\vspace{.5em}
The selection problem has become more difficult because model repositories have grown rapidly. Platforms such as Hugging Face host a large and continuously changing collection of models contributed by researchers, companies, and community members. This openness accelerates innovation and reuse, but it also creates a practical decision problem for software engineers who must identify models that fit a concrete application context~\cite{faulconbridge2023professionals}. A practitioner may need to compare different model families, fine-tuned variants, quantized versions, licenses, hardware requirements, and quality-related trade-offs before making an informed choice.

\vspace{.5em}
Relevant evidence is distributed across heterogeneous and unevenly maintained sources. Repository metadata may describe task labels, licenses, model size, and download counts; model cards and README files may describe intended use, datasets, limitations, and capabilities; community discussions may report practical experiences with speed, stability, integration difficulty, or unexpected behavior. These sources differ in completeness, terminology, and reliability. Consequently, foundation-model selection is a cognitively demanding software-engineering decision that requires more than simple keyword search or popularity-based filtering. We argue that foundation-model selection must be treated as an explicit and auditable software-component selection decision, supported by systematic evidence synthesis and transparent multi-criteria decision models.

\subsection*{Limitations of existing approaches}

Existing approaches to foundation-model discovery and selection can be broadly categorized into three groups: popularity and metadata-based filtering, keyword-based retrieval methods, and large language model (LLM)-based recommendation systems.

\vspace{.5em}
Popularity and metadata-based approaches rely on indicators such as download counts, likes, and manually defined tags to recommend models. While these signals facilitate basic discovery, they provide only surface-level representations of model characteristics and rarely capture task-specific requirements. For example, knowledge graph-based approaches that leverage Hugging Face metadata improve the structural relationships between models and tasks, yet remain constrained by incomplete and inconsistently defined metadata \cite{chen2025benchmarking}.

\vspace{.5em}
Keyword-based retrieval methods attempt to match user queries with textual descriptions of models. These approaches partially address the challenge of heterogeneous information sources by incorporating unstructured text such as documentation or scientific literature. However, they rely primarily on lexical matching and therefore lack semantic understanding of relationships between features and task requirements. As a result, they struggle to differentiate between models with similar descriptions but different functional capabilities \cite{di2024automated}.

\vspace{.5em}
More recently, LLM-based recommendation systems have been explored for model selection tasks. These approaches can interpret complex user queries and reason about multiple criteria simultaneously, enabling more context-aware recommendations. Nevertheless, their reasoning processes remain largely opaque due to the black-box nature of LLM architectures. Additionally, they are susceptible to hallucinations and may produce inaccurate or nonexistent recommendations, thereby limiting their reliability in decision-support scenarios \cite{zhao2024recommender}.

\subsection*{Research gap}

Despite the growing need for effective foundation-model selection mechanisms, existing approaches address only isolated aspects of the decision problem. Popularity and metadata-based filters are scalable but shallow; keyword-based retrieval can surface relevant model-card text but does not explicitly model trade-offs; and conversational LLM recommenders can produce fluent advice but do not provide stable, auditable, criterion-level ranking logic. Current approaches therefore provide limited support when practitioners must compare many alternatives under context-specific functional and quality constraints.

\vspace{.5em}
Data-driven decision frameworks have been successfully applied to other technology-selection domains, including software packages, AI models, blockchain platforms, and blockchain oracle selection~\cite{farshidi2018decision,farshidi2020decision,farshidi2021decision,ahmadjee2025decision,farshidi2025empirical,joonbakhsh2025evidence}. Recent MCDM studies also confirm the value of structured decision support for AI-related choices, including ML reproducibility assessment~\cite{math14091536}, healthcare deep learning model evaluation~\cite{Drissi2024multi}, trustworthy AI application assessment~\cite{Alsalem2024evolution}, MLaaS cloud selection~\cite{BHOL2024909}, AI decision support~\cite{OLABANJO2026100723}, and fuzzy software selection~\cite{Seker2021Pythagorean}. However, these studies usually evaluate small, fixed sets of alternatives using predefined expert criteria. They do not address repository-scale foundation-model ecosystems where criteria and evidence must be derived automatically from incomplete metadata, model descriptions, and community feedback.

\vspace{.5em}
This creates a gap for explainable, repository-scale decision support that treats foundation-model selection as an auditable software-component selection decision involving functional requirements, non-functional quality concerns, and operational constraints. Such support should not only recommend models, but also expose why a model is recommended, which criteria contribute to its score, and which trade-offs remain for the practitioner to inspect.

\subsection*{Research Questions}
This work addresses the following research questions:

\begin{itemize}
  \item \textbf{RQ1:} How accurately can functional features and perceived quality attributes be extracted from unstructured repository content and community discussions at scale?
  \item \textbf{RQ2:} How does an explainable multi-criteria decision framework compare with practical baseline selection methods in terms of model-level and family-level recommendation quality?
  \item \textbf{RQ3:} How do practitioners perceive the usability, transparency, and usefulness of HugSelect's explainable decision-support framework?
\end{itemize}

\subsection*{Solution overview (HugSelect)}

To address the identified research gap, this study proposes \textit{HugSelect}, a data-driven decision-support framework for foundation-model selection based on Multi-Criteria Decision Making (MCDM)~\cite{triantaphyllou2000multi}. HugSelect operationalizes foundation-model selection as a single coherent decision-support method that integrates automated evidence extraction with explicit multi-criteria ranking. Each candidate model is assessed against functional requirements, operational constraints, and community-perceived quality attributes using a weighted decision model, making the selection rationale inspectable and auditable.

\vspace{.5em}
HugSelect integrates heterogeneous sources of model-related information into a unified knowledge base. Automated data pipelines process repository metadata, model-card descriptions, and community feedback using feature extraction, sentiment analysis, and quality-attribute mapping techniques. The resulting knowledge base represents model alternatives together with metadata features, functional capabilities, perceived quality signals, and traceability links to supporting evidence.

\vspace{.5em}
Users interact with the system through a natural-language interface. An LLM-based intent extractor translates a user query into structured requirements and candidate criteria. These criteria are mapped to the knowledge base and evaluated using a Weighted Sum Model (WSM / SAW). Rather than relying on a mathematical innovation in decision theory, the framework adapts WSM/SAW specifically to maximize ranking transparency and scalability. Instead of returning only a natural-language answer, HugSelect produces a ranked list of candidate models and decomposes each recommendation into criterion-level score contributions, enabling users to inspect trade-offs and adjust priorities.

\vspace{.5em}
Compared with metadata-based filtering, HugSelect provides richer representations of model capabilities. Compared with keyword-based retrieval, it explicitly models trade-offs among criteria. Compared with LLM-only recommendation, it provides more transparent and controllable ranking logic. To support open science and reproducibility, the complete source code and accompanying datasets are publicly available~\cite{adali2026huggingface}.

\subsection*{Contributions}

This study makes the following contributions:

\begin{itemize}
\item \textbf{A repository-scale decision-support method} that formulates foundation-model selection as an explicit and auditable software-component selection decision.
\item \textbf{An integrated evidence-synthesis pipeline} that automatically harvests repository metadata, model-card descriptions, functional capabilities, and community-derived perceived quality signals into a unified knowledge base of 71,274 Hugging Face models.
\item \textbf{An explainable WSM/SAW-based decision engine} that maps user requirements to structured criteria and decomposes recommendation scores into inspectable criterion-level contributions.
\item \textbf{An empirical evaluation} comprising extraction pipeline validation, comparative retrieval analysis against commercial LLM baselines over 44 literature-derived scenarios, fine-grained ablation experiments, and an exploratory TAM-based practitioner user study.
\end{itemize}

\subsection*{Evaluation Caveats and Scope}
To support proper interpretation of our findings, three methodological boundaries should be noted early: (i) the 44 selection scenarios utilize peer-reviewed literature model choices as proxy ground truth rather than single absolute benchmark targets; (ii) commercial LLM baselines reflect zero-shot, time-bound conversational system outputs; and (iii) the 10-participant user study provides exploratory usability and acceptance evidence based on the Technology Acceptance Model (TAM) rather than a comparative trial of decision quality.

\subsection*{Paper organization}
Section~\ref{sec:related} reviews related work. Section~\ref{sec:framework} presents the HugSelect framework and methodology. Section~\ref{sec:implementation} describes the system implementation. Section~\ref{sec:evaluation} reports the evaluation. Section~\ref{sec:discussion} discusses implications and limitations. Section ~\ref{sec:conclusion} concludes.

\section{Related Work}
\label{sec:related}

foundation-model selection sits at the intersection of software engineering, machine learning, repository mining, recommender systems, and decision science. To make the positioning easier to follow, we group the literature around three questions: (i) what kinds of model repositories and alternatives have been studied, (ii) what evidence is used to characterize models, and (iii) how candidate models are compared and evaluated. Appendix~\ref{app:gap-analysis} provides the full gap-analysis table used to derive this positioning.

\begin{table}
\centering
\scriptsize

\setlength{\tabcolsep}{4pt}
\renewcommand{\arraystretch}{1.12}
\caption{Summary of related-work gaps addressed by HugSelect.}
\label{tab:related_work_summary}
\begin{tabularx}{\linewidth}{p{2.8cm} X X}
\toprule
\textbf{Theme} & \textbf{Main focus in prior work} & \textbf{Remaining gap addressed by HugSelect} \\
\midrule
Repository scale and scope & Prior studies often analyze small sets of alternatives, specific model families, or repository metadata snapshots. Large-scale Hugging Face studies exist, but usually focus on classification, tracing, or metadata-based discovery. & Large-scale selection support is still limited when alternatives are numerous, heterogeneous, and continuously changing. HugSelect analyzes a curated set of 71,274 Hugging Face models with sufficient metadata and textual evidence. \\
\addlinespace
Evidence sources & Existing approaches typically rely on popularity signals, tags, model cards, benchmark results, or manually defined criteria. & Functional capabilities and perceived quality signals from community discussions are rarely integrated into a unified decision model. HugSelect combines metadata, model descriptions, and community feedback. \\
\addlinespace
Decision support & LLM-based and keyword-based systems can produce fluent recommendations, while MCDM studies provide transparent comparison logic in smaller decision settings. & There is limited support for transparent multi-criteria trade-off reasoning over large foundation-model repositories. HugSelect uses WSM/SAW to produce auditable criterion-level rankings. \\
\addlinespace
Evaluation & Prior work commonly evaluates classification, retrieval, benchmark performance, or user perception in isolation. & Few studies jointly validate extraction quality, recommendation quality, and perceived usefulness. HugSelect combines pipeline validation, comparative case studies, and a practitioner user study. \\
\bottomrule
\end{tabularx}
\end{table}

\subsection{Repository Scale and Scope}

foundation-model repositories have grown rapidly in recent years~\cite{bommasani2021opportunities,awais2025foundation}. Hugging Face has become a central platform for open model sharing and discovery, but its scale also makes manual selection difficult. Existing studies vary substantially in scope. Some surveys focus on general foundation-models or specific technical areas such as vision-language models, large language models, multimodal models, agents, or medical imaging~\cite{chen2024evolution,azad2023foundational,zhang2021language,ding2024learning,minaee2024large,mienye2025large,zhou2024taxonomy}. Other work focuses on software engineering use cases or on the role of pre-trained models in software development~\cite{faulconbridge2023professionals,gonzalez2025pre}.

Most selection-oriented studies evaluate relatively small sets of alternatives. Exceptions include large-scale repository-mining studies such as those by ~\cite{cao2021dekr}, ~\cite{liu2023task}, and ~\cite{gonzalez2025pre}. The most closely related large-scale Hugging Face work is the knowledge-graph-based study by ~\cite{chen2025benchmarking}, which supports recommendation, classification, and tracing over a large Hugging Face graph. However, that work does not focus on multi-criteria trade-off reasoning or on quality evidence derived from community discussions. ~\cite{di2024automated} and ~\cite{suryani2024exploration} also study Hugging Face models, but their focus is categorization and cross-repository exploration rather than explainable model selection.

\subsection{Evidence Sources and Model Characterization}

Model selection depends on more than task labels or popularity. Practitioners often need to understand functional capabilities, supported modalities, architecture families, license constraints, deployment requirements, and quality concerns. Prior work uses several evidence sources, including academic literature, GitHub repositories, model hubs, public datasets, and benchmarks~\cite{jain2022hugging,gonzalez2025pre,di2024automated,suryani2024exploration}. SearchSECO is particularly relevant as an example of ecosystem-scale indexing, where heterogeneous software artifacts are collected, represented, and searched across repository boundaries~\cite{jansen2020searchseco}. Repository mining, app-store mining, and software-ecosystem indexing research shows that large-scale software artifacts and unstructured text from documentation, issue trackers, Q\&A sites, and user reviews can be transformed into structured indicators such as provenance links, topics, quality attributes, and defect-related signals~\cite{martin2017survey,jansen2020searchseco}.

Existing model-selection approaches usually use only a subset of these signals. Metadata extraction is scalable but shallow; benchmark evidence is useful but expensive and incomplete across large repositories; and manual expert criteria provide structure but do not scale well to thousands of alternatives. HugSelect therefore integrates three complementary evidence families: structured repository metadata, functional features extracted from model descriptions, and perceived perceived quality attributes derived from community feedback. This design follows earlier data-driven decision frameworks for software packages, AI models, and technology selection~\cite{farshidi2018decision,farshidi2020decision,farshidi2021decision,farshidi2025empirical,joonbakhsh2025evidence}, but adapts them to the larger and more text-heavy setting of foundation-model repositories.

\subsection{Decision-Making and Evaluation Strategies}

Several decision-making strategies have been used for model and software-artifact recommendation. AI/ML-based methods infer suitability from usage data or model characteristics~\cite{chen2025benchmarking,ding2024learning}. Knowledge-based and expert-driven systems rely on manually encoded relationships. Data-driven recommender systems use interaction data, collaborative filtering, or ranking models~\cite{ricci2010introduction,lu2012recommender,adomavicius2010multi}. MCDM methods, such as AHP, TOPSIS, PROMETHEE, WSM, and fuzzy variants, support explicit comparison across multiple criteria~\cite{triantaphyllou2000multi,vaidya2006analytic,behzadian2012state,chakrabortty2023multi,radulescu2025criteria}.

Recent MCDM studies confirm the value of structured decision support for AI- and software-related choices, including ML reproducibility readiness, healthcare model evaluation, trustworthy AI assessment, MLaaS selection, blockchain oracle selection, AI decision support, and fuzzy software selection~\cite{math14091536,Drissi2024multi,Alsalem2024evolution,BHOL2024909,ahmadjee2025decision,OLABANJO2026100723,Seker2021Pythagorean}. However, these studies typically evaluate small, fixed sets of alternatives using predefined expert criteria. They do not address large-scale foundation-model repositories, where criteria and evidence must be derived automatically from incomplete metadata, model descriptions, and community discussions.

\subsection{Positioning of HugSelect}

HugSelect differs from prior work in three ways. First, it targets repository-scale foundation-model selection rather than small fixed alternative sets. Second, it combines structured metadata with textual functional features and community-derived quality indicators. Third, it uses an explicit WSM/SAW-based MCDM model so that recommendations can be decomposed into criterion-level contributions. This choice prioritizes transparency, low elicitation burden, and efficient ranking over methodological complexity. More advanced MCDM methods can be integrated in future versions, but WSM/SAW is appropriate for the first scalable implementation because users can adjust priorities without pairwise comparisons, fuzzy membership functions, or complex preference calibration.

The full comparison matrix in Appendix~\ref{app:gap-analysis} shows that existing studies rarely combine large-scale automated data collection, multi-source feature extraction, community-feedback-based quality mapping, MCDM-based ranking, and practitioner evaluation in a single framework. HugSelect is designed to fill this gap.

\section{The HugSelect Framework (Approach)}
\label{sec:framework}

This section presents the conceptual and methodological foundations of HugSelect, grounded in Design Science Research (DSR) methodology \cite{hevner2004design}. The goal is to clarify the problem formulation, architectural structure, data integration strategy, decision-making mechanism, and explainability design, independently of any particular implementation.

\subsection{Research Methodology}
\label{sec:methodology}

The development of HugSelect follows the DSR methodology, which focuses on solving real-world problems through the systematic design, development, and evaluation of artifacts \cite{hevner2004design}. DSR is particularly well-suited to this work because it enables a practical, rigorously evaluated solution to a complex, real-world decision-making problem. The research process comprises the following phases:

\begin{enumerate}
  \item \textbf{Problem identification:} Identifying gaps and limitations in existing foundation-model selection approaches, particularly the reliance on simple metadata signals and the lack of explainability.
  \item \textbf{Literature review:} Conducting a systematic literature study using snowballing \cite{wohlin2014guidelines} to inform design decisions and establish novelty. The review covers related domains including LLM selection, machine learning model selection, and intelligent agent selection.
  \item \textbf{Framework design:} Developing a theoretical framework encompassing data collection and integration pipelines, knowledge base construction, and decision model formulation.
  \item \textbf{Artifact construction:} Implementing the HugSelect decision support system as a concrete instantiation of the framework.
  \item \textbf{Evaluation:} Assessing the validity of feature extraction pipelines and the overall effectiveness of HugSelect through empirical validation, comparative case studies, and user studies.
\end{enumerate}

The literature study provides a stable basis for the design decisions of the adapted framework, supported by existing academic work, informing choices regarding information sources, feature extraction methods, knowledge base formation, and decision-making criteria and strategies.

\subsection{Problem Formulation and Unit of Analysis}
\label{sec:problem}

Let $M = \{m_1, \dots, m_n\}$ be the set of available model alternatives. 

\paragraph{Unit of Analysis.} In HugSelect, the unit of analysis $m_i \in M$ is an individual repository artifact hosted on Hugging Face. The indexed collection of 71,274 models includes base foundation-models (e.g., \texttt{meta-llama/Meta-Llama-3-8B}), fine-tuned checkpoints (e.g., \texttt{meta-llama/Meta-Llama-3-8B-Instruct}), quantized variants (e.g., \texttt{TheBloke/Llama-2-7B-GGUF}), and adapter weights. Individual model entries retain their specific metadata and operational characteristics (such as parameter size, license, and format) while being mapped to their broader architectural model family in the knowledge base.

\paragraph{Feature Families vs. Decision Criteria.} Each model $m_i$ is characterized by a feature vector $\mathbf{f}_i = (f_{i1}, \dots, f_{ik})$ across $k$ individual decision criteria $c_j$. These criteria belong to three feature families:
\begin{itemize}
  \item \textbf{Metadata features ($F_M$):} Structured repository attributes (e.g., task tag, license, parameter size, download count).
  \item \textbf{Functional features ($F_F$):} Capability descriptors extracted from model cards and README documentation (e.g., reasoning capabilities, domain specificity, supported modalities).
  \item \textbf{Perceived quality attributes ($F_Q$):} Non-functional signals derived from community feedback, structured according to ISO/IEC 25010 (e.g., perceived reliability, perceived performance efficiency).
\end{itemize}

A user provides a selection context $q$ consisting of a natural-language description of application requirements and constraints. An intent extraction component transforms $q$ into structured criteria weights $\mathbf{w} = (w_1, \dots, w_k)$ where $w_j \ge 0$ and $\sum_{j=1}^k w_j = 1$.

The framework computes a ranking score $R(m_i, q)$ for each candidate model using a Weighted Sum Model (WSM / SAW)~\cite{triantaphyllou2000multi}:
\begin{equation}
  R(m_i, q) = \sum_{j=1}^{k} w_j \cdot s_{ij}
  \label{eq:wsm}
\end{equation}
where $s_{ij} \in [0,1]$ is the normalized score of model $m_i$ on criterion $c_j \in F_M \cup F_F \cup F_Q$. The decision goal is to generate a prioritized ordering of $M$ and expose the criterion-level contributions $w_j \cdot s_{ij}$ for auditability.

\paragraph{Running Example.} To illustrate the framework operations throughout Sections~\ref{sec:framework} and \ref{sec:implementation}, we trace a single running scenario: \textit{"A software engineering team requires an open-weights foundation-model to power a clinical Q\&A assistant. Contextual constraints require: task = Text-Generation, modality = Text, license = Permissive (Apache-2.0 or MIT), deployment format = Quantized GGUF ($<10$GB), with strong community-perceived performance efficiency and reliability."}

\subsection{Conceptual Architecture and Workflow}
\label{sec:architecture}

The architectural design of HugSelect is adapted from the multi-criteria decision-making framework proposed by Farshidi~\cite{farshidi2020multi}. As illustrated in Figure~\ref{fig:framework}, the adapted framework enables automated data collection, handling of heterogeneous information sources, systematic mapping of extracted features, and linking the knowledge base to an MCDM structure for decision-making. The framework comprises four interconnected layers:

\begin{figure}
  \centering
  \includegraphics[width=1.0\linewidth]{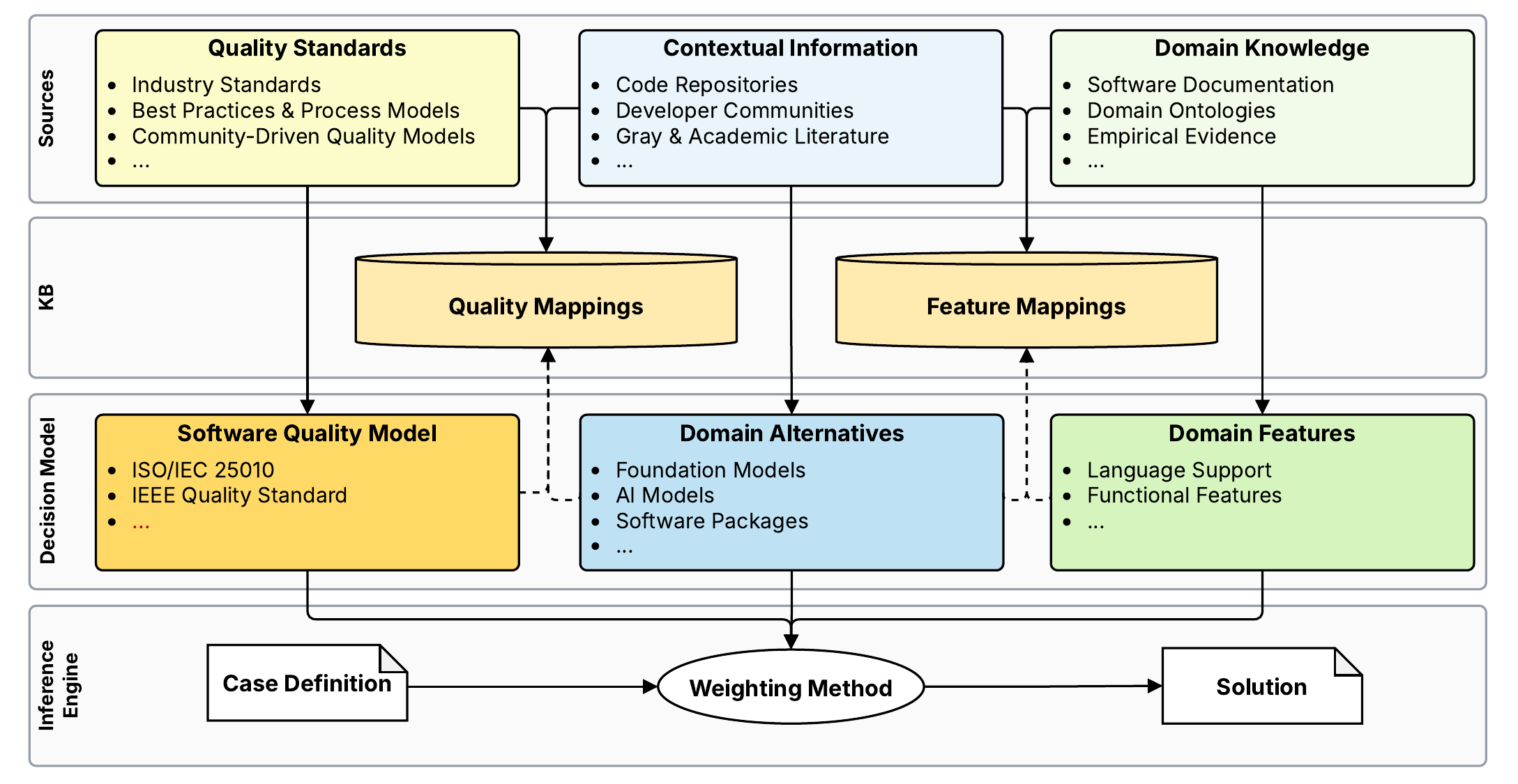}
  \caption{Adapted multi-layer framework architecture for foundation-model selection (adapted from \cite{farshidi2020multi}). The framework integrates heterogeneous data sources into a unified knowledge base and connects it to an MCDM-based inference engine.}
  \label{fig:framework}
\end{figure}

\begin{enumerate}
  \item \textbf{Source Layer:} Ingests heterogeneous evidence from model hubs (Hugging Face metadata and model cards) and developer discussion platforms (Reddit, Stack Overflow).
  \item \textbf{Knowledge Base Layer:} Integrates extracted metadata, functional capability phrases, and community-perceived quality signals into a graph-structured schema, preserving traceability to original source snippets.
  \item \textbf{Decision Model Layer:} Processes natural-language queries to map user intent into candidate criteria across the three feature families and assign initial criteria weights.
  \item \textbf{Inference Engine Layer:} Executes criterion score normalization, applies WSM ranking (Equation~\ref{eq:wsm}), and decomposes scores into explainable criterion-level contributions.
\end{enumerate}

The architecture is intentionally modular, allowing different extraction techniques, decision models, or data sources to be substituted without structural changes. This modularity supports extensibility and adaptation to evolving model ecosystems.

\subsection{Conceptual Model}
\label{sec:conceptual}

\begin{figure}
  \centering
  \includegraphics[width=0.9\linewidth]{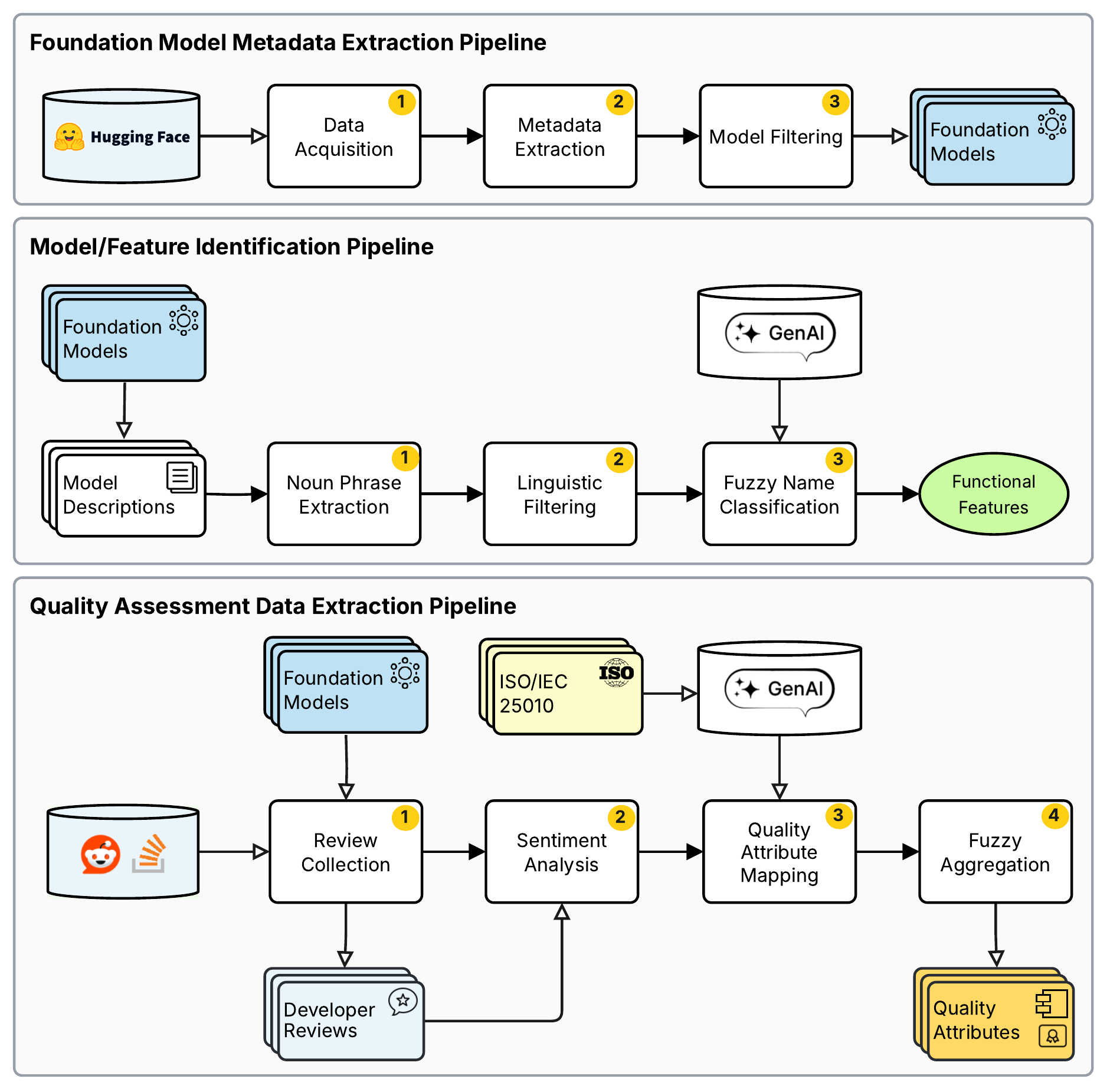}
  \caption{Conceptual model of the HugSelect framework showing the data collection pipelines, feature extraction, and knowledge base construction.}
  \label{fig:conceptual}
\end{figure}

Figure~\ref{fig:conceptual} presents the conceptual model of HugSelect, illustrating how the framework integrates metadata, functional features, and quality attributes into a unified knowledge base to enable model selection. The conceptual workflow proceeds as follows:

\begin{enumerate}
  \item \textbf{Data Collection:} Three pipelines harvest repository metadata, model descriptions (from model cards and README files), and community feedback (user comments, reviews, and Q\&A posts).
  \item \textbf{Feature Extraction:} Unstructured text from model descriptions and community feedback is transformed into structured functional features and quality attributes through dedicated extraction pipelines.
  \item \textbf{Knowledge Base Integration:} All extracted features are mapped to a unified schema that links models to their metadata, capabilities, quality dimensions, and supporting evidence, enabling traceability across heterogeneous sources.
  \item \textbf{Query Processing:} User queries are analyzed to determine relevant criteria and weights, which are then used to instantiate the decision model.
  \item \textbf{Ranking and Explanation:} The MCDM engine computes model rankings and generates explanations that decompose scores into criterion-level contributions.
\end{enumerate}

This conceptual model bridges the abstract architectural layers with the practical application of HugSelect, clarifying how heterogeneous information sources are systematically transformed into actionable selection recommendations.

\subsection{Data Integration Methodology}
\label{sec:integration}

HugSelect synthesizes three complementary evidence sources to construct a holistic representation of each foundation-model:

\paragraph{Repository Metadata ($F_M$).} Structured fields (task tags, framework, license, downloads, creation date) provide baseline operational boundaries. Missing or inconsistent tags are flagged and handled via schema mapping. In our running example, metadata filters immediately identify models offering GGUF files and permissive licensing.

\paragraph{Model Card Descriptions ($F_F$).} Unstructured README text and model cards provide narrative details on capabilities, training data, and intended use. Dependency parsing extracts technical capability phrases, mapping them to standardized functional descriptors.

\paragraph{Community Feedback ($F_Q$).}
User comments, reviews, and Q\&A posts related to specific models are collected from the hub and external platforms. After cleaning and model-level linking, they are processed by the quality attribute extraction pipeline to yield scores along selected quality dimensions.

The integration methodology defines a unified schema that stores all three data types as attributes of model entities, mapping rules from raw text to standardized functional and quality dimensions, and consistency checks to ensure that extracted features are plausible given the underlying metadata.

This multi-source integration strategy addresses the heterogeneity challenge inherent in foundation-model ecosystems, where relevant information is distributed across structured metadata, unstructured documentation, and community-generated content.

\subsection{Multi-Criteria Decision Making}
\label{sec:mcdm}

The MCDM component operationalizes the problem formulation by implementing a systematic approach to criterion definition, normalization, weighting, and ranking \cite{triantaphyllou2000multi}. Its responsibilities are:

\paragraph{1. Criterion Definition \& Intent Mapping.} For query $q$, the framework identifies relevant criteria $c_j$ across $F_M, F_F, F_Q$. In our running example, intent extraction identifies 5 criteria: $c_1=\text{Task: Text-Generation}$ ($F_F$), $c_2=\text{License: Permissive}$ ($F_M$), $c_3=\text{Format: GGUF}$ ($F_M$), $c_4=\text{Perceived Efficiency}$ ($F_Q$), and $c_5=\text{Perceived Reliability}$ ($F_Q$).

\paragraph{2. Normalization.} Heterogeneous criterion values $f_{ij}$ are normalized to scores $s_{ij} \in [0,1]$. Categorical matches (e.g., license, task) assign binary scores $s_{ij} \in \{0, 1\}$, continuous metrics (e.g., parameter size, downloads) use min-max scaling, and textual/quality attributes assign similarity or aggregated fuzzy quality values.

\paragraph{3. Weighting \& WSM Ranking.} Weights $w_j$ are assigned based on user constraints (e.g., $w_1=0.25, w_2=0.25, w_3=0.20, w_4=0.15, w_5=0.15$). The engine evaluates Equation~\ref{eq:wsm} for all candidate models. HugSelect uses weighted sum as the default because of its simplicity and interpretability, but the architecture permits alternative methods such as TOPSIS or PROMETHEE when needed.

\paragraph{4. Sensitivity Analysis.} The engine computes ranking variations under weight perturbations ($\pm 20\%$), informing users whether top recommendations are sensitive to specific preference trade-offs.

\subsection{Explainability Design}
\label{sec:explainability}

To ensure complete transparency, HugSelect generates explanations along three dimensions:
\begin{itemize}
  \item \textbf{Score Decomposition:} Exposes the exact breakdown of $R(m_i, q)$ into criterion-level contributions $w_j \cdot s_{ij}$, allowing users to see why a candidate model ranks highest.
  \item \textbf{Feature-Level Traceability:} Links scores back to source evidence (e.g., highlighting exact model card sentences or community review snippets).
  \item \textbf{Comparative Trade-off Views:} Renders side-by-side criteria comparisons across top candidates, illustrating trade-offs between model variants.
\end{itemize}

\section{System Implementation}
\label{sec:implementation}

This section describes the technical implementation of the HugSelect framework, detailing its architecture, data processing pipelines, multi-criteria decision-making engine, and user interface components.

\subsection{Implementation Architecture}
\label{subsec:architecture}

The HugSelect framework is implemented as a four-layer system designed to automate foundation-model selection from the Hugging Face Hub. The architecture comprises:

\begin{enumerate}
    \item \textbf{Automated Data Collection Layer:} Interfaces with the Hugging Face, Reddit, and Stack Exchange APIs to gather model metadata, community discussions, and developer feedback.
    
    \item \textbf{Knowledge Base Layer:} Stores structured representations of models, their functional features, quality attributes, and relationships in a graph-based format.
    
    \item \textbf{LLM-based Intent Detection Layer:} Processes natural language queries to extract user requirements and map them to structured decision criteria.
    
    \item \textbf{MCDM Engine Layer:} Implements multi-criteria ranking algorithms to generate ranked model recommendations with transparent rationale.
\end{enumerate}


\subsection{Data Collection and Processing}
\label{subsec:data_processing}

\paragraph{Data Sources and Collection Pipeline}
The data collection pipeline interfaces with three primary APIs. The Hugging Face API~\cite{huggingface_api} retrieves model metadata, including name, task type, architecture family, download counts, license information, and publication dates, resulting in an initial collection of approximately 71,274 models. The Reddit API~\cite{reddit_api} collects community discussions, user experiences, and informal reviews from machine learning subreddits. The StackExchange API~\cite{stackexchange_api} collects technical questions, answers, and developer feedback on specific models. 

\vspace{.5em}
Data quality filtering removes duplicates, deprecated models, and entries with incomplete metadata to ensure knowledge base integrity. The complete open-source implementation of these data collection and processing pipelines along with the initial raw data and the outputs of our functional and quality-attribute processing pipelines are systematically archived in our Mendeley repository \cite{adali2026huggingface}.

\paragraph{Functional Feature Extraction Pipeline}
Functional features describe the technical capabilities and intended use cases of foundation-models. The extraction pipeline employs noun-phrase filtering with spaCy's dependency parser~\cite{spacy} to identify technical noun phrases in model descriptions and documentation. Feature clustering then groups extracted features into three primary dimensions: modality (categorizing models by input/output types such as text, image, audio, multimodal, with 11 distinct labels in the knowledge base), task (classifying models by problem type such as text generation, classification, translation, with 52 task categories in the dataset), and family (grouping models sharing architectural lineage or design principles such as LLaMA, BERT, Stable Diffusion, with 87 model families tracked). Because developer documentation is often incomplete, the system employs simple information-extraction methods to classify model features based on available descriptions. Models that cannot be definitively assigned to clusters are retained in the knowledge base but receive lower recommendation priority. Table~\ref{tab:unified_model_summary} summarizes the distribution of models across the modality, task, and family dimensions, showing the top 5 entries in each category.

\begin{table}
\scriptsize
\centering
\begin{tabular}{lllr}
\hline
\textbf{Category} & \textbf{\#Items }& \textbf{Item} &\textbf{ Count} \\
\hline

\multirow{6}{*}{Modality} & \multirow{6}{*}{11}
& Text & 46,489 \\
& & Multimodal & 14,470 \\
& & Image & 4,356 \\
& & Audio & 3,438 \\
& & Reinforcement Learning & 473 \\
& & Other / Unclear & 1,735 \\

\hline
\multirow{6}{*}{Task} & \multirow{6}{*}{52}
& Text-Generation & 35,556 \\
& & Text-to-Image & 11,579 \\
& & Text-Classification & 3,788 \\
& & Automatic-Speech-Recognition & 2,241 \\
& & Image-Classification & 1,862 \\
& & Other / Unclear & 1,735 \\

\hline
\multirow{6}{*}{Family Root} & \multirow{6}{*}{87}
& LLaMA & 14,789 \\
& & BERT & 8,237 \\
& & Stable Diffusion & 7,819 \\
& & Mistral & 7,194 \\
& & Qwen & 4,697 \\
& & Other / Unclear & 8,259 \\

\hline
\multicolumn{3}{l}{\textbf{Total Models}} & \textbf{71,274} \\
\hline
\end{tabular}
\caption{Top 5 entries per category. In total, the dataset contains 11 modality labels, 52 task categories, and 87 model families. Only the most frequent entries are shown for readability, with remaining entries grouped as Other / Unclear where applicable.}
\label{tab:unified_model_summary}
\end{table}

\subsubsection{Quality Attribute Extraction Pipeline}

Quality attributes in HugSelect represent community-perceived non-functional characteristics of foundation-models, based on feedback collected from Hugging Face discussions, Reddit, and Stack Overflow. Because queries based only on model names can introduce entity ambiguity and unrelated discussions, the pipeline applies several filtering stages before quality mapping. First, only review snippets that explicitly mention the model are retained. Second, an LLM-based relevance check removes cases in which the matched name refers to a different entity rather than to a foundation-model. Third, code-heavy or structurally noisy snippets are removed during preprocessing. The remaining reviews are analyzed with three multilingual transformer-based sentiment models, and neutral reviews are discarded because they do not provide clear evidence about perceived model quality.

\vspace{.5em}
The filtered review snippets are then mapped to ISO/IEC 25010-inspired quality characteristics, including functional suitability, performance efficiency, reliability, interaction capability, maintainability, security, flexibility, and safety. Mapping is performed through LLM prompting that interprets semantic cues in the review text and assigns the most relevant quality dimension. For example, statements about slow inference or high resource consumption correspond to performance efficiency, comments about crashes or inconsistent outputs indicate reliability, and remarks about documentation or integration difficulty relate to maintainability or interaction capability. Examples of such mappings are shown in Table~\ref{tab:quality_mapping_examples}.

\vspace{.5em}
Quality characteristics are defined in accordance with ISO/IEC 25010, while community reviews provide observable indicators of perceived quality rather than controlled measurements of runtime behavior. In this study, the mapping from review statements to quality attributes is treated as a coding task that operationalizes these constructs. Review snippets serve as qualitative evidence of user-perceived system behavior and can indicate properties such as performance efficiency, reliability, or maintainability. The LLM is therefore used as a coding instrument that assigns review statements to predefined quality categories based on their semantic content.

\vspace{.5em}
To reduce subjectivity when translating informal feedback into formal quality attributes, the pipeline requires that each review snippet be categorized along with its sentiment and a brief rationale for the classification. Rather than relying on a single observation, quality signals are aggregated across multiple independent reviews.

To ensure robustness, a quality characteristic is evaluated for a model only when supported by at least three evidence-bearing (positive or negative) review instances ($L + H \ge 3$). Neutral reviews are discarded during preprocessing as they do not provide explicit directional evidence regarding perceived quality. This sufficiency threshold reduces the influence of isolated or ambiguous comments.

\paragraph{Fuzzy Aggregation Scoring.} For eligible models ($L + H \ge 3$), the review evidence is aggregated into a normalized perceived quality score $S_{fuzzy}(m_i, Q_k) \in [0, 1]$ computed as the ratio of positive reviews over all directional evidence:
\begin{equation}
  S_{fuzzy}(m_i, Q_k) = \frac{H}{L + H}
  \label{eq:fuzzy_aggregation}
\end{equation}
where $H$ represents the count of positive review snippets and $L$ represents the count of negative review snippets mapped to quality characteristic $Q_k$. If a model has insufficient community evidence ($L + H < 3$), no score is assigned for that attribute due to data sparsity, ensuring newly published or sparsely reviewed models are not unfairly evaluated on unrepresentative feedback.

\begin{table}
\scriptsize

\centering
\caption{Examples of mapping review snippets to ISO/IEC 25010-inspired quality characteristics.}
\label{tab:quality_mapping_examples}
\begin{tabular}{p{5.2cm} p{3cm} p{1.8cm} p{4.5cm}}
\hline
\textbf{Review Snippet} & \textbf{Quality Attribute} & \textbf{Sentiment} & \textbf{Rationale} \\
\hline
``DeepSeek-R1's innovative text generation and reasoning features would be a perfect fit for our project.'' & Functional Suitability & Positive & Indicates that the model effectively fulfills the intended task requirements. \\

``The storytelling capabilities of GPT-2 are incredible.'' & Functional Suitability & Positive & Highlights strong task performance in text generation. \\

``I tried 10 to 15 different settings, and it still didn't work.'' & Reliability & Negative & Describes repeated failure of the model to perform correctly. \\

``The model cannot be found even though it appears in the Hugging Face model list.'' & Reliability & Negative & Indicates issues with availability or stability during use. \\

``I loaded my SentenceTransformer model locally and successfully embedded sample sentences.'' & Functional Suitability & Positive & Confirms correct operation of the model's primary embedding function. \\
\hline
\end{tabular}
\end{table}

\subsection{Knowledge Base Structure}
\label{subsec:knowledge_base}

The knowledge base represents foundation-models and their attributes as a graph-structured integration layer, enabling flexible querying, unified access to heterogeneous evidence, and traceability between models, features, quality attributes, and their source data. In this work, the knowledge graph is used primarily as an integration and traceability layer rather than as an autonomous graph-reasoning engine. Its role is to unify heterogeneous evidence from metadata, model descriptions, and community feedback, while the ranking logic is implemented in the MCDM decision engine.

\vspace{.5em}
Figure~\ref{fig:kg_sample} illustrates an example knowledge graph comparing three variations of the DeepSeek-R1 model~\cite{deepseek_r1}: the base model \texttt{deepseek-ai/DeepSeek-R1}, a quantized variant \texttt{unsloth/DeepSeek-R1-GGUF}, and an activation-aware quantized variant \texttt{cognitivecomputations/DeepSeek-R1-AWQ}. The graph captures three types of relationships. Metadata relations link models to their technical specifications, although these are not depicted in the simplified visualization. Functional features establish many-to-many relationships between models and extracted capabilities derived from NLP analysis of descriptions. Quality attributes connect models to community-derived indicators inferred from reviews and discussions.

\definecolor{goldorange}{HTML}{FFD966}

\begin{figure}
\centering
\scriptsize
\begin{tikzpicture}[
    model/.style={draw, rounded corners, align=center, minimum width=2.5cm, minimum height=0.8cm, fill=gray!10},
    feature/.style={draw, rounded corners, align=center, minimum width=2.3cm, minimum height=0.65cm, fill=green!10},
    quality/.style={draw, diamond, aspect=2.2, align=center, inner sep=1pt, fill=goldorange},
    rel/.style={-{Latex[length=2mm]}, thin},
    every node/.style={font=\scriptsize}
]

\node[model] (base) {DeepSeek-R1\\base model};
\node[model, below left=1.5cm and 0.2cm of base] (gguf) {DeepSeek-R1-GGUF\\quantized variant};
\node[model, below right=1.5cm and 0.2cm of base] (awq) {DeepSeek-R1-AWQ\\AWQ variant};

\node[feature, above left=1.2cm and 0.1cm of base] (reasoning) {reasoning};
\node[feature, above=1.2cm of base] (textgen) {text generation};
\node[feature, above right=1.2cm and 0.1cm of base] (api) {API compatibility};

\node[feature, below=1.1cm of gguf] (quant) {quantization};
\node[feature, below=1.1cm of awq] (resource) {resource efficiency};

\node[quality, left=2.7cm of base] (fs) {functional\\suitability};
\node[quality, right=2.7cm of base] (perf) {performance\\efficiency};
\node[quality, below=2.9cm of base] (relia) {reliability};

\draw[rel] (base) -- (reasoning);
\draw[rel] (base) -- (textgen);
\draw[rel] (base) -- (api);
\draw[rel] (gguf) -- (quant);
\draw[rel] (awq) -- (resource);
\draw[rel] (gguf) -- (base) node[midway, left, sloped] {variant of};
\draw[rel] (awq) -- (base) node[midway, right, sloped] {variant of};
\draw[rel, dashed] (fs) -- (base);
\draw[rel, dashed] (perf) -- (gguf);
\draw[rel, dashed] (perf) -- (awq);
\draw[rel, dashed] (relia) -- (base);

\end{tikzpicture}
\caption{Simplified knowledge-graph example for three DeepSeek-R1 variants. Rectangles denote model entities and functional features; diamonds denote perceived quality attributes derived from community feedback. The full knowledge base contains many more models, features, and evidence links, but this simplified view highlights the relation types used for matching and explanation.}
\label{fig:kg_sample}
\end{figure}
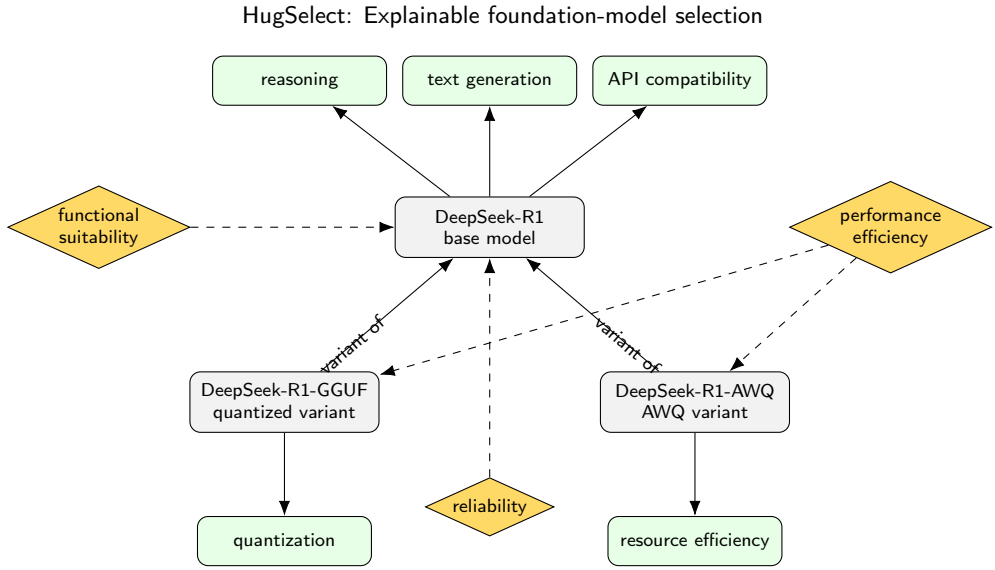

\vspace{.5em}
This graph-based representation supports semantic matching by organizing model variants, functional features, and quality attributes in a common structure. The matching and ranking are performed by the decision engine, while the knowledge base provides the integrated evidence needed to compare model variations and near-substitutes when they satisfy the same user intent. This distinction improves the validity of evaluation by separating semantic correctness from exact model-ID matching. The comprehensive set of model clusters, functional features, and quality attributes mapped within this knowledge base is available for review in our data repository \cite{adali2026huggingface}.

\subsection{MCDM Engine Implementation}
\label{subsec:mcdm_engine}

The MCDM engine is implemented as a modular Python library that orchestrates the decision-making process from query processing to ranked output generation.

\subsubsection{User Intent Extraction}

\begin{figure}
    \centering
        \includegraphics[width=0.8\linewidth]{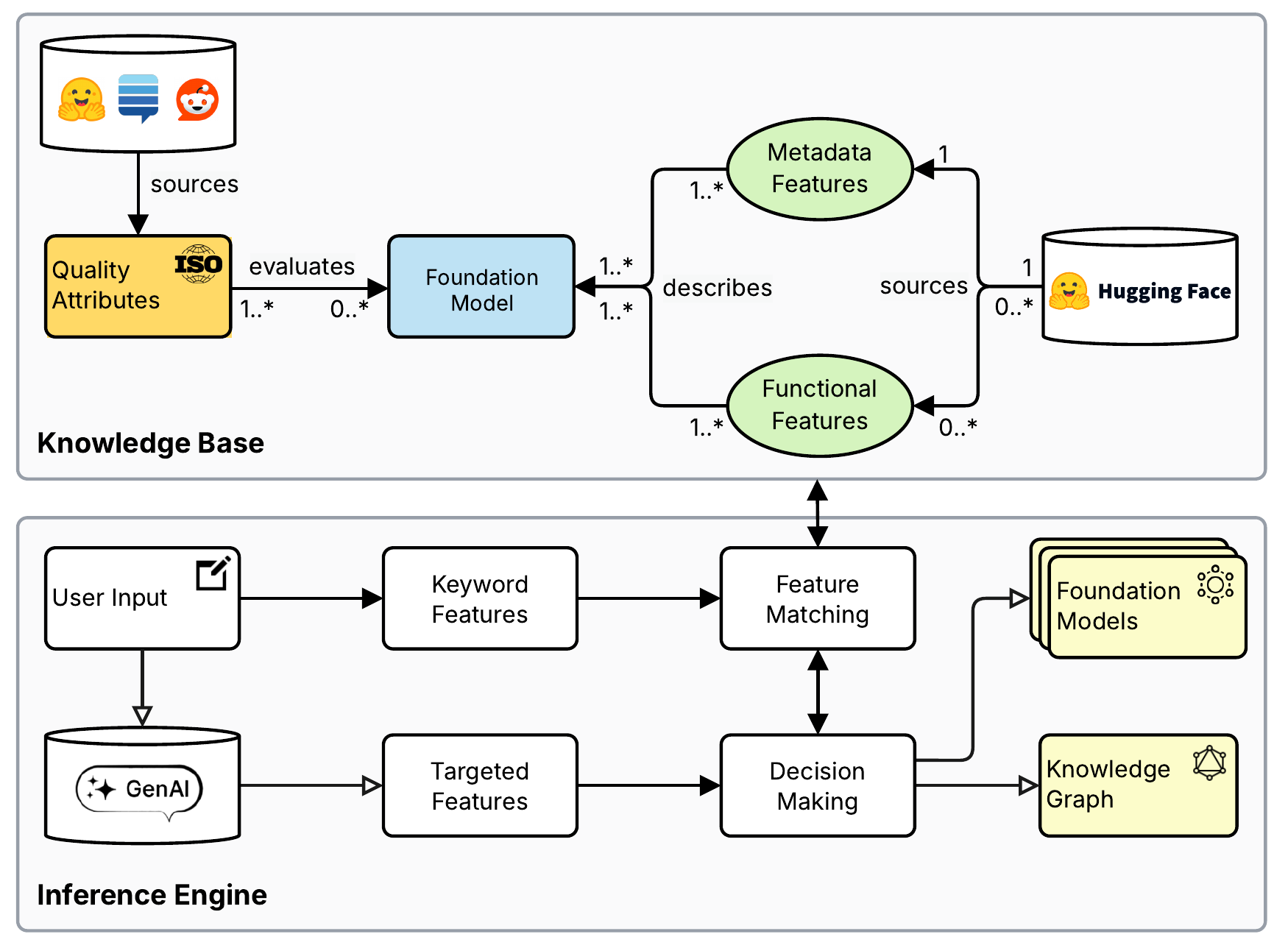}
        \caption{Inference engine of the HugSelect framework. User queries are converted into structured features and matched against a knowledge graph of foundation-models to generate ranked model recommendations.}\label{fig:inference_kg}
\end{figure}

Figure~\ref{fig:inference_kg} illustrates the inference architecture. When users submit natural-language queries describing their intended task, application context, and operational constraints, the system processes them using two parallel extractors. The keyword feature extractor identifies explicit technical terms and requirements using pattern matching and domain-specific lexicons. The LLM-based targeted feature extractor uses LLaMA~\cite{llama} and Gemini~\cite{gemini} variants to interpret implicit requirements and contextual constraints. This component addresses potential biases by using multiple models and aggregating their outputs.

The extracted features are mapped to the structured feature space of the knowledge base, creating a user requirement vector aligned with model attributes. This design is related to earlier context-aware search systems in computational environments, such as CANSF, which derives search queries from a user's active Jupyter notebook context and retrieves semantically related notebooks~\cite{li2022contextaware}.

\subsubsection{Weighted Criteria Vector and MoSCoW Prioritization}

User requirements are transformed into a weighted criteria vector following the MoSCoW prioritization method~\cite{clegg1994case}:

\begin{itemize}
    \item \textbf{Must-have (M):} Critical requirements treated as hard constraints (restriction properties).
    \item \textbf{Should-have (S):} Important preferences with high weights.
    \item \textbf{Could-have (C):} Desirable features with moderate weights.
    \item \textbf{Won't-have (W):} Explicitly excluded features.
\end{itemize}

The LLM-based extractor assigns importance weights to each identified feature based on linguistic cues in the user query (e.g., "must support," "preferably," "ideally").

\subsubsection{Candidate Retrieval and Matching}

The inference engine compares the user requirement vector against model feature vectors in the knowledge base. The matching process incorporates:

\begin{enumerate}
    \item \textbf{Degree of Match:} Cosine similarity between user requirements and model features for functional attributes; threshold-based matching for categorical metadata.
    
    \item \textbf{Feature Importance:} Weights from the MoSCoW prioritization amplify the contribution of critical features to the overall score.
    
    \item \textbf{Popularity Metrics:} Download counts and community engagement serve as tie-breakers when multiple models have similar functional matches.
\end{enumerate}

\subsubsection{Normalization and Ranking}

Retrieved candidates undergo normalization to ensure comparability across heterogeneous criteria:

\begin{itemize}
    \item \textbf{Min-Max Normalization:} Applied to quantitative metrics (downloads, model size) to scale values to $[0,1]$.
    \item \textbf{Binary Normalization:} Applied to categorical features (license type, modality) using exact- or partial-match scoring.
\end{itemize}

The final suitability score for each model $m$ is computed as:

$$
\text{Score}(m) = \sum_{i=1}^{n} w_i \cdot \text{match}_i(m, u)
$$

where $w_i$ is the importance weight of criterion $i$, $\text{match}_i(m, u)$ is the normalized match score between model $m$ and user requirement $u$ on criterion $i$, and $n$ is the total number of criteria.

Models are ranked in descending order of their suitability scores, producing a prioritized candidate list.

\subsubsection{Explainable Output Generation}

The engine generates transparent rationale for each recommendation through three mechanisms. Score decomposition breaks down the total score into contributions from metadata, functional features, and quality attributes, enabling users to understand the relative importance of each dimension. Feature highlighting identifies which user requirements were satisfied by each recommended model and which were not, providing clear justification for ranking decisions. Knowledge graph visualization presents contextual information about model relationships and feature overlaps in an interactive graph format, allowing users to explore alternative models and understand trade-offs. This explainability layer supports informed decision-making and builds user trust in the recommendations.

\subsection{End-to-End Selection Workflow}
\label{subsec:end_to_end_workflow}

To clarify how the implementation operationalizes the framework, Table~\ref{tab:end_to_end_example} presents an illustrative end-to-end workflow. The example shows how a natural-language request is transformed into structured criteria, how candidate models are retrieved from the knowledge base, and how the final ranking is explained. The example is not intended as a new benchmark case; rather, it demonstrates the decision-support logic implemented by HugSelect.

\begin{table}
\centering
\scriptsize

\setlength{\tabcolsep}{5pt}
\renewcommand{\arraystretch}{1.15}
\caption{Illustrative end-to-end workflow in HugSelect.}
\label{tab:end_to_end_example}
\begin{tabularx}{\linewidth}{p{3.2cm} X}
\toprule
\textbf{Step} & \textbf{Example operation} \\
\midrule
User request & A practitioner asks for a foundation-model for reasoning-heavy question answering, with a preference for text generation, efficient deployment, permissive licensing, and reliable community feedback. \\
Requirement extraction & The intent extractor identifies task-related criteria (reasoning, question answering, text generation), operational preferences (efficient deployment, model size or quantization), and restriction properties (license compatibility). \\
Criteria weighting & Critical requirements are treated as must-have constraints, while preferred capabilities and perceived quality attributes receive lower but explicit weights using the MoSCoW-inspired weighting scheme. \\
Candidate retrieval & The knowledge base retrieves models whose metadata and functional features match the task and modality requirements. Related variants, such as base, quantized, or fine-tuned versions, remain comparable through family-level mappings. \\
Ranking & The MCDM engine normalizes heterogeneous criteria and computes a weighted suitability score for each candidate model. \\
Explanation & The output decomposes the score into functional, metadata, and perceived quality contributions and highlights which requirements are satisfied, partially satisfied, or unsupported by available evidence. \\
Decision support & The practitioner can inspect trade-offs, adjust weights, compare related variants, and use the ranking as decision support rather than as an automatic final decision. \\
\bottomrule
\end{tabularx}
\end{table}

This workflow illustrates the intended use of HugSelect as an auditable selection aid. The system does not claim that the highest-ranked model is universally optimal; instead, it makes the selection rationale explicit so that practitioners can examine whether the ranking aligns with their project constraints and risk tolerance.

\section{Evaluation}
\label{sec:evaluation}

This section presents an empirical evaluation of the HugSelect framework through three complementary approaches: pipeline validation, comparison with commercial LLM-based baselines, and a user study. The evaluation addresses the three research questions outlined in Section~\ref{sec:introduction} and follows Design Science Research (DSR) principles for artifact validation.

\vspace{.5em}
To ensure full transparency and enable independent verification, all data used and generated throughout these evaluation phases are publicly available in our replication package \cite{adali2026huggingface}. The repository is organized to reflect our methodology, containing dedicated records for the pipeline validation (functional features and quality attributes), the detailed case study execution (including curation papers, recommendation results, and ablation studies), and the complete empirical results from the user study.

\subsection{Evaluation Overview}
\label{subsec:eval_overview}

The evaluation strategy comprises three phases:

\begin{enumerate}
    \item \textbf{Pipeline Validation (RQ1):} Assesses the accuracy and reliability of the functional feature extraction pipeline (P2) and the quality attribute extraction pipeline (P3) through ground-truth comparison and expert validation.
    
    \item \textbf{Framework Validation via Case Studies (RQ2):} Evaluates HugSelect's recommendation performance against four commercial LLM-based baselines (ChatGPT, Claude, Gemini, Perplexity) across 44 real-world model-selection scenarios, using both model-level and family-level metrics.
    
    \item \textbf{User Study (RQ3):} Examines user perceptions of recommendation quality, transparency, and usability through subjective assessments and task performance measurements.
\end{enumerate}

\subsection{Pipeline Validation (RQ1)}

To assess the reliability of the automated extraction pipelines, we validated both functional feature extraction (P2) and quality attribute extraction (P3) against reference datasets constructed through multi-LLM consensus and human annotation. Validation used a stratified random sample of 500 models from the 71,274-model knowledge base (250 from the top 10,000 by popularity, 250 from the remainder) and 500 community reviews. Metrics included precision, recall, F1-score, and Intersection over Union (IoU) for feature extraction, and accuracy with macro- and weighted-precision, recall, and F1 for sentiment and quality-attribute classification.

\subsubsection{Functional Feature Extraction (P2)}

To construct a reference dataset, three LLMs (Gemini, GPT, LLaMA) independently extracted functional features from model descriptions. Features that appeared in at least two LLM outputs were retained and validated by a human expert, thereby reducing single-model bias.

\vspace{.5em}
To reduce circularity in evaluating the functional feature extraction pipeline, we constructed the reference ground truth through a human-in-the-loop validation process rather than relying solely on LLM-generated outputs. For the sampled models, three LLMs independently extracted candidate functional features from the model descriptions, while a domain expert manually annotated functional features using the same criteria. A feature was included in the ground-truth set only when it was identified by the human annotator and confirmed by at least one LLM, ensuring that retained features were both human-validated and reproducible from the source text. This protocol uses the domain expert as the primary filter for factual and contextual relevance, while the LLMs provide an additional consistency check across independent semantic interpretations. Therefore, the resulting ground truth supports evaluation of HugSelect’s extraction quality without depending exclusively on automated model agreement.

\vspace{.5em}
Table~\ref{tab:functional_feature_comparison} presents validation results against each individual LLM reference set and the unified ground truth.

\begin{table}
\centering
\scriptsize
\begin{tabular}{lcccc}
\toprule
\textbf{Source} & \textbf{Precision} & \textbf{Recall} & \textbf{F1} & \textbf{IoU} \\
\midrule
Gemini   & 0.802 & 0.778 & 0.790 & 0.653 \\
LLaMA    & 0.834 & 0.582 & 0.686 & 0.522 \\
ChatGPT  & 0.692 & 0.677 & 0.684 & 0.520 \\
\midrule
\textbf{Ground truth}  & \textbf{0.746} & \textbf{0.865} & \textbf{0.801} & \textbf{0.668} \\
\bottomrule
\end{tabular}
\caption{Functional feature extraction evaluation results.}
\label{tab:functional_feature_comparison}
\end{table}

\vspace{.5em}
The pipeline achieved strong recall (0.865) and F1-score (0.801) against the unified reference, with IoU of 0.668, indicating reliable capture of model capabilities. Gemini showed the highest agreement with the ground truth. Manual error analysis of 100 mismatched features revealed that false positives primarily stemmed from training metadata (11 cases), infrastructure details (13 cases), and dataset references (6 cases), while false negatives arose from specialized deployment terminology (14 cases), optimization techniques (18 cases), and task-specific architectural descriptions (16 cases). Table~\ref{tab:fp_fn_error_examples} provides representative examples. These errors reflect the heterogeneity of developer-written model cards, which range from marketing-focused narratives to benchmark-heavy technical reports. Filtering contextual information that does not directly describe functional capabilities remains a challenge for automated extraction.

\begin{table}
\centering
\scriptsize
\begin{tabular}{lll}
\toprule
\textbf{Category (Count)} & \textbf{Type} & \textbf{Example} \\
\midrule
\multirow{3}{*}{Training metadata (11)}
& FP & fine tuned model \\
& FP & base language model \\
& FP & large language model \\

\multirow{3}{*}{Infrastructure details (13)}
& FP & CUDA \\
& FP & google colab \\
& FP & normalization \\

\multirow{3}{*}{Dataset references (6)}
& FP & conll2003 dataset \\
& FP & russian text \\
& FP & italian \\

\midrule

\multirow{3}{*}{Deployment formats (14)}
& FN & GGUF \\
& FN & Quantized Model \\
& FN & Weight-only quantization \\

\multirow{3}{*}{Optimization techniques (18)}
& FN & QLoRA \\
& FN & lr\_scheduler\_warmup\_ratio \\
& FN & positional encoding method \\

\multirow{3}{*}{Task descriptions (16)}
& FN & latent text-to-image diffusion model \\
& FN & SQL statement \\
& FN & future n-gram prediction \\
\bottomrule
\end{tabular}
\caption{Representative false-positive (FP) and false-negative (FN) extraction errors.}
\label{tab:fp_fn_error_examples}
\end{table}

\subsubsection{Quality Attribute Extraction (P3)}

Quality attribute extraction operates on community reviews filtered for relevance and sentiment polarity. In the available evaluation dataset, the sentiment-analysis stage covers 51,352 filtered review snippets. Of these, 17,867 were labelled positive, 28,233 neutral, and 5,252 negative. Because neutral snippets do not provide clear evidence about perceived quality, the quality-mapping stage focuses on the remaining evidence-bearing snippets and assigns 30,214 review instances to ISO/IEC 25010-inspired categories, including unclear cases. We validated both sentiment classification and ISO/IEC 25010 quality-attribute mapping using 500 annotated reviews. Three LLMs independently labeled each review; labels appearing in at least two outputs formed the reference set, which was then refined by a human annotator. Table~\ref{tab:combined_sentiment_quality_results} reports classification performance, and Table~\ref{tab:coverage_combined} summarizes both the annotated sample and the full filtered dataset.

\begin{table}
\centering
\scriptsize
\setlength{\tabcolsep}{4pt}
\renewcommand{\arraystretch}{1.1}
\begin{tabularx}{\linewidth}{l l *{7}{>{\centering\arraybackslash}X}}
\toprule
\textbf{Task} & \textbf{Source} & \textbf{Accuracy} &
\multicolumn{3}{c}{\textbf{Macro}} &
\multicolumn{3}{c}{\textbf{Weighted}} \\
\cmidrule(lr){4-6} \cmidrule(lr){7-9}
 &  &  & Precision & Recall & F1 & Precision & Recall & F1 \\
\midrule
\multirow{3}{*}{Sentiment}
 & ChatGPT            & 0.65 & 0.64 & 0.61 & 0.59 & 0.73 & 0.65 & 0.66 \\
 & Gemini             & 0.63 & 0.63 & 0.60 & 0.57 & 0.72 & 0.63 & 0.65 \\
 & \textbf{Ground Truth} & \textbf{0.70} & \textbf{0.66} & \textbf{0.70} & \textbf{0.63} 
                         & \textbf{0.79} & \textbf{0.70} & \textbf{0.72} \\
\addlinespace
\multirow{3}{*}{Quality Mapping}
 & ChatGPT            & 0.77 & 0.59 & 0.52 & 0.53 & 0.79 & 0.77 & 0.76 \\
 & Gemini             & 0.72 & 0.56 & 0.49 & 0.49 & 0.80 & 0.72 & 0.73 \\
 & \textbf{Ground Truth} & \textbf{0.84} & \textbf{0.68} & \textbf{0.63} & \textbf{0.65} 
                         & \textbf{0.84} & \textbf{0.84} & \textbf{0.83} \\
\bottomrule
\end{tabularx}
\caption{Sentiment analysis and quality-attribute mapping performance.}
\label{tab:combined_sentiment_quality_results}
\end{table}

\begin{figure}
\centering
\includegraphics[width=0.8\linewidth]{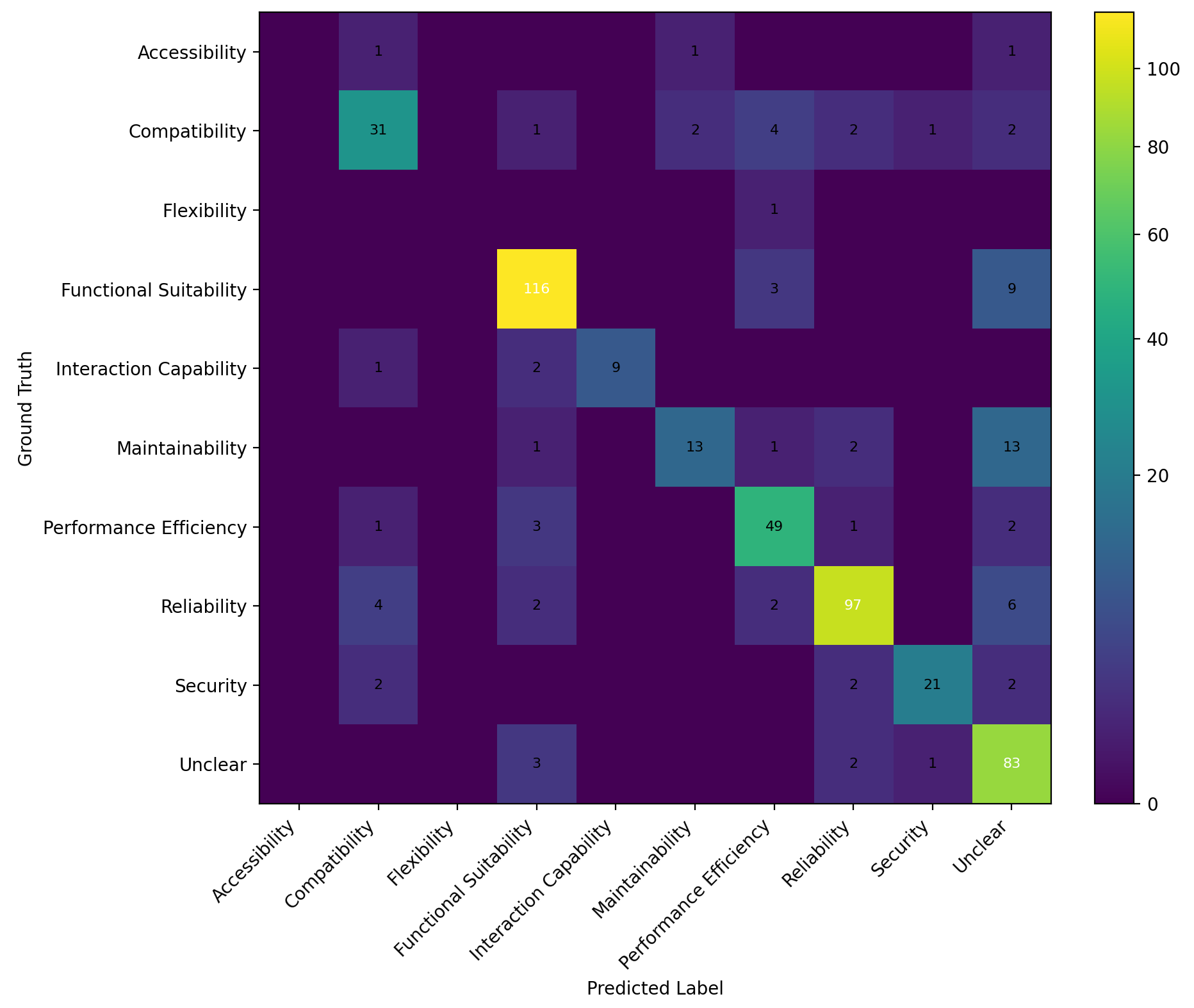}
\caption{Quality attribute mapping confusion matrix.}
\label{fig:confusion_quality}
\end{figure}

\vspace{.5em}
The pipeline achieved accuracy of 0.70 for sentiment classification and 0.84 for quality attribute mapping. Weighted metrics exceeded macro metrics due to class imbalance; neutral-sentiment reviews (55\% of the full dataset) and unclear quality attributes (23.74\%) were the most frequent. The confusion matrix (Figure~\ref{fig:confusion_quality}) shows that unclear attributes were most often misclassified, reflecting the difficulty of distinguishing vague opinions from specific quality concerns in code-heavy, technical discussions. Filtering neutral reviews before quality mapping improved precision by removing low-information content, though separating subjective opinions from technical discourse remains challenging when reviews mix debugging notes, code snippets, and evaluative statements.

Table~\ref{tab:coverage_combined} summarizes the distribution of sentiment and quality attribute labels in both the annotated sample and the full filtered dataset, illustrating the prevalence of functional suitability concerns and the relative scarcity of safety-related feedback.

\begin{table}
\centering
\scriptsize
\caption{Sentiment and quality attribute label distribution.}
\label{tab:coverage_combined}
\begin{tabular}{llrrrr}
\toprule
\textbf{Task} & \textbf{Category} &
\multicolumn{2}{c}{\textbf{Sample}} &
\multicolumn{2}{c}{\textbf{Full Dataset}} \\
\cmidrule(lr){3-4} \cmidrule(lr){5-6}
 &  & Count & (\%) & Reviews & (\%) \\
\midrule
\multirow{3}{*}{Sentiment Analysis}
 & Positive & 135 & 27.0 & 17867 & 34.8 \\
 & Neutral  & 255 & 51.0 & 28233 & 55.0 \\
 & Negative & 110 & 22.0 & 5252  & 10.2 \\
\midrule
\textbf{Total} &  & \textbf{500} & \textbf{100} & \textbf{51352} & \textbf{100} \\
\midrule
\multirow{10}{*}{Quality Mapping}
 & Functional Suitability & 119 & 23.8 & 8806 & 29.15 \\
 & Unclear                & 97  & 19.4 & 7173 & 23.74 \\
 & Reliability            & 66  & 13.2 & 4821 & 15.96 \\
 & Compatibility          & 54  & 10.8 & 3983 & 13.18 \\
 & Performance Efficiency & 44  & 8.8  & 3248 & 10.75 \\
 & Security               & 30  & 6.0  & 472  & 1.56 \\
 & Interaction Capability & 30  & 6.0  & 382  & 1.26 \\
 & Maintainability        & 30  & 6.0  & 1245 & 4.12 \\
 & Flexibility            & 30  & 6.0  & 46   & 0.15 \\
 & Safety                 & 0   & 0.0  & 10   & 0.03 \\
\midrule
\textbf{Total} &  & \textbf{500} & \textbf{100} & \textbf{30214} & \textbf{100} \\
\bottomrule
\end{tabular}
\end{table}

\textbf{Summary for RQ1 (Pipeline Validation):} Automated extraction pipelines achieve high accuracy at repository scale, yielding an F1-score of 0.80 (IoU=0.67) for functional capability features and 0.84 accuracy for mapping community review snippets to ISO/IEC 25010 perceived quality attributes.

\subsection{Framework Validation via Case Studies (RQ2)}
\label{subsec:case_study}

\subsubsection{Experimental Design}

We conducted a comparative evaluation of HugSelect against four commercial LLM-based recommendation systems across 44 literature-derived model selection scenarios.

\paragraph{Baselines and Reproducibility}
We compared HugSelect against four practical LLM-based recommendation baselines: ChatGPT, Claude, Gemini, and Perplexity. The baselines were evaluated as zero-shot conversational recommenders using the same case descriptions. Each system was asked to return a ranked list of candidate foundation-models for the same selection scenario. The goal was not to claim a definitive benchmark of the underlying proprietary models, but to compare HugSelect with the type of recommendation support practitioners can obtain from widely used conversational systems. Because provider-managed interfaces can change over time and do not always expose decoding parameters, the comparison should be interpreted as a time-bound practical baseline comparison rather than a fully controlled model-to-model benchmark. Table~\ref{tab:baseline_reproducibility} summarizes the configuration information recorded for the evaluation.

\begin{table}
\centering
\scriptsize
\setlength{\tabcolsep}{4pt}
\renewcommand{\arraystretch}{1.12}
\caption{Baseline configuration information used in the comparative evaluation.}
\label{tab:baseline_reproducibility}
\begin{tabularx}{\linewidth}{p{2.2cm} p{3.2cm} p{2.4cm} X}
\toprule
\textbf{System} & \textbf{Interface / model family} & \textbf{Mode} & \textbf{Recorded evaluation setting} \\
\midrule
HugSelect & Local implementation with the curated knowledge base & Deterministic ranking & Same 44 case descriptions; top-10 ranked recommendations; WSM/SAW scoring with criterion-level explanation. \\
ChatGPT & GPT-4-based conversational recommender & Zero-shot & Same 44 case descriptions; top-10 candidate models requested; no fine-tuning or prompt optimization. Provider decoding settings were not exposed. \\
Claude & Claude 3 Opus conversational recommender & Zero-shot & Same 44 case descriptions; top-10 candidate models requested; no fine-tuning or prompt optimization. Provider decoding settings were not exposed. \\
Gemini & Gemini Pro conversational recommender & Zero-shot & Same 44 case descriptions; top-10 candidate models requested; no fine-tuning or prompt optimization. Provider decoding settings were not exposed. \\
Perplexity & Perplexity AI with web-search augmentation & Zero-shot & Same 44 case descriptions; top-10 candidate models requested; no fine-tuning or prompt optimization. Provider decoding settings were not exposed. \\
\bottomrule
\end{tabularx}
\end{table}

\paragraph{Case Selection}
The 44 cases were systematically sampled to cover different modalities, task types, and application contexts. For each case, we recorded contextual attributes such as domain, modality, task, and selected model. Table~\ref{tab:case_study_coverage} summarizes this distribution. Because some cases belong to more than one domain or task category, category counts should be interpreted as multi-label descriptors rather than mutually exclusive totals. This coverage is important because foundation-model selection depends on interactions among domain characteristics, task requirements, and modality constraints. At the model level, the dataset includes 32 unique models, and the most frequent model appears in only four cases, reducing the risk that the evaluation is dominated by a small set of popular models.

\paragraph{Ground Truth Construction}
The evaluation uses literature-derived proxy ground truth based on expert-informed model selections reported in peer-reviewed scientific studies. For each case, ground truth was established through a three-step process. First, candidate scientific papers were collected from the Semantic Scholar and OpenAlex databases using systematic keyword searches targeting studies that reported the use of Hugging Face models. Second, associated GitHub repositories mentioned in the abstracts were examined to verify explicit references to the reported models, and only papers from established conferences or journals were retained. Third, the remaining papers were manually reviewed to include only application-oriented cases in which a foundation-model was explicitly selected for a defined task, resulting in a curated set of scientific cases whose reported model choices were used as proxy ground truth for evaluation. Recommendations of each system were evaluated according to whether they recovered the reference model or its corresponding model family; other suitable alternatives may exist. Therefore, the case-study labels should be interpreted as proxy ground truth for comparative retrieval analysis, not as proof that the literature-reported model is the only correct or optimal choice.

\begin{table}
\centering
\scriptsize
\setlength{\tabcolsep}{6pt}
\renewcommand{\arraystretch}{1.15}
\begin{tabularx}{\linewidth}{l >{\raggedright\arraybackslash}X c}
\hline
\textbf{Type} & \textbf{Category} & \textbf{Count} \\
\hline

\multirow{7}{*}{\textbf{Domain}} 
& Language \& Communication   & 14 \\
& Education \& Knowledge      & 6  \\
& Healthcare \& Medicine      & 5  \\
& Media \& Creative           & 10 \\
& Security \& Ethics          & 3 \\
& Human-Computer Interaction  & 5  \\
& Science \& Methodology      & 6  \\
\hline

\multirow{4}{*}{\textbf{Modality}} 
& Text / NLP        & 22 \\
& Computer Vision   & 8  \\
& Multimodal        & 14 \\
\hline

\multirow{8}{*}{\textbf{Task}} 
& Text Generation              & 7  \\
& Summarization                & 6  \\
& Image Generation             & 5  \\
& Image Captioning \& Grounding& 3  \\
& Text Classification          & 5  \\
& Visual Understanding         & 6  \\
& Question Answering           & 10 \\
& Translation                  & 2  \\
\hline

\textbf{Total Cases} & & \textbf{44} \\
\hline

\end{tabularx}
\caption{Distribution of domains, modalities, and tasks across the 44 evaluation cases. Counts are multi-label descriptors and therefore may not sum to the total number of cases within each group.}
\label{tab:case_study_coverage}
\end{table}

\subsubsection{Evaluation Metrics}

We employed two evaluation levels to capture complementary aspects of recommendation quality.

\paragraph{Model-Level (ML) Metrics}
Model-level metrics evaluate exact matches of model identifiers. \textbf{Coverage@10} measures whether at least one reference model appears among the top-10 recommendations for a case. We also report \textbf{Overlap@10 with HugSelect} (Overlap@10) for the baseline systems as a diagnostic agreement measure, indicating the number of baseline recommendations that overlap with HugSelect's top-10 list. This overlap measure is not a relevance metric; therefore, HugSelect itself is marked as the reference system in Table~\ref{tab:retrieval_results_extended}.

\paragraph{Family-Level (FL) Metrics}
Family-level metrics evaluate model family matches (e.g., BERT variants or GPT variants), treating family members as potential substitutes when they satisfy the same selection intent. \textbf{Coverage@10} measures whether the relevant model family is represented in the top-10 recommendations. \textbf{Precision@10} is computed as the mean proportion of top-10 recommended families that belong to the expanded relevant family set for each case. This expanded set can contain multiple acceptable families when the source study, associated repository, or expert review indicates that more than one family is a valid substitute. \textbf{NDCG@10} measures ranking quality with position-weighted relevance:
\[
\text{NDCG@}K = \frac{\text{DCG@}K}{\text{IDCG@}K}
\]
where higher-ranked relevant items contribute more to the score.

Family-level metrics are particularly important for practical model selection because users often consider multiple variants within a model family as viable alternatives. The distinction between exact model identifiers and family-level substitutes also reduces the risk of penalizing semantically valid recommendations that differ only in version, quantization, or fine-tuning variant.

\subsubsection{Statistical Significance and Effect Sizes}

An omnibus \textbf{Friedman test} on family-level rank distributions across all systems showed no significant global difference ($\chi^2 = 8.786$, $p = 0.0667$). Pairwise \textbf{Wilcoxon signed-rank tests} between HugSelect and commercial LLMs confirmed non-significant variations ($p > 0.18$ across all pairs). McNemar tests at the model level showed HugSelect significantly outperformed Gemini ($p < 0.05$, $OR = 6.00$) and metadata/filtering baselines ($p < 0.001$), while performing comparably to ChatGPT and Claude.

\subsubsection{Results}

Table~\ref{tab:retrieval_results_extended} presents the full empirical comparison.

\begin{table}
\centering
\caption{Performance comparison of HugSelect against baseline systems (n=44 cases) at cutoff 10 across Model-Level (ML) and Family-Level (FL). Overlap@10 reports agreement with HugSelect's top-10 list. Statistical significance relative to HugSelect is assessed via McNemar test; * $p<0.05$, *** $p<0.001$.}
\label{tab:retrieval_results_extended}
\begin{tabular}{lcccccc}
\toprule
\multirow{2}{*}{\textbf{System}} & \multicolumn{2}{c}{\textbf{Model-Level}} & \multicolumn{4}{c}{\textbf{Family-Level}} \\
\cmidrule(lr){2-3} \cmidrule(lr){4-7}
& \textbf{Cov@10} & \textbf{Ovlp@10} & \textbf{Cov@10} & \textbf{Ovlp@10} & \textbf{Prec@10} & \textbf{NDCG@10} \\
\midrule
\textbf{HugSelect} & \textbf{0.61} & - & \textbf{0.91} & - & \textbf{0.37} & \textbf{0.74} \\
ChatGPT & 0.52 & 0.12 & 0.86 & 0.36 & 0.41 & 0.72 \\
Claude & 0.64 & 0.12 & 0.93 & 0.45 & 0.43 & 0.79 \\
Gemini & 0.39* & 0.06 & 0.89 & 0.37 & 0.29 & 0.69 \\
Perplexity & 0.52 & 0.10 & 0.84 & 0.38 & 0.29 & 0.67 \\
\bottomrule
\end{tabular}
\end{table}

\vspace{.5em}
HugSelect achieved coverage competitive with the strongest LLM baselines at both the model level (0.61, second only to Claude's 0.64) and family level (0.91, close to Claude's 0.93 and Gemini's 0.89). At the model level, HugSelect achieved higher coverage than Gemini and Perplexity, and the McNemar test showed a statistically significant difference only against Gemini ($p<0.05$). Against Claude, the observed difference was small and not statistically significant. These findings indicate that HugSelect is competitive with commercial conversational systems on retrieval effectiveness while providing structured, criterion-level transparency. The NDCG@10 score of 0.74 places HugSelect among the stronger systems in terms of ranking quality, close to ChatGPT (0.72) and below Claude (0.79). The gap between model-level (0.61) and family-level (0.91) coverage suggests that HugSelect often surfaces relevant model families even when exact model identifiers differ, which is useful in settings where multiple variants may satisfy the same intent. The modest Precision@10 (0.37), in line with all baselines (0.29--0.43), reflects the exploratory nature of model selection, where users may benefit from diverse recommendations spanning multiple families rather than a narrow list of near-duplicates.

\paragraph{Qualitative Analysis}

Manual inspection of recommendation outputs revealed systematic differences between HugSelect and the LLM baselines. HugSelect provided structured recommendations with explicit feature matching, quality scores, and knowledge graph context, enabling users to trace why each model was recommended. In contrast, LLM baselines generated natural language explanations but often recommended popular models (e.g., BERT, GPT-2) regardless of specific requirements, exhibiting recency and popularity bias. While their explanations were fluent, they were sometimes factually inconsistent with the model's actual capabilities.

\vspace{.5em}
Figure~\ref{fig:kg_sample} illustrates how HugSelect's knowledge graph representation enables transparent comparison of model variations, supporting informed decision-making.

\subsection{User Study}

Ten participants with domain knowledge in AI, computer science, and data science evaluated HugSelect. After a brief orientation, participants explored the system for ten minutes and completed a structured survey using a 7-point Likert scale (1 = Strongly Agree, 7 = Strongly Disagree). Responses were analyzed using descriptive statistics and Cronbach's alpha for internal consistency. Given the small sample size, reliability values are interpreted as descriptive indicators rather than definitive psychometric evidence.

\vspace{.5em}
The survey was based on the Technology Acceptance Model (TAM) \cite{davis1989perceived} and TAM 2 \cite{venkatesh2000theoretical}. TAM posits that system usage is determined by perceived usefulness (PU) and perceived ease of use (PEOU), which influence attitude toward using (ATT) and behavioral intention to use (ITU). TAM 2 introduced five external variables: subjective norm (SN), image (IMG), job relevance (JR), output quality (OQ), and result demonstrability (RD). The survey included four statements for each core construct and two for each external variable.

\subsubsection{Results}

Table~\ref{tab:tam-survey} presents response distributions, mean scores, standard deviations, and Cronbach's alpha for each construct.

\begin{table}
\centering
\scriptsize 
\setlength{\tabcolsep}{8pt} 
\begin{tabular}{lccccccc|ccc}
\hline
\textbf{Construct} 
& \multicolumn{7}{c|}{\textbf{Responses}} 
& \multicolumn{3}{c}{\textbf{Statistics}} \\
\cline{2-11}
& \rotatebox{70}{SA (1)}
& \rotatebox{70}{A (2)}
& \rotatebox{70}{SLA (3)}
& \rotatebox{70}{N (4)}
& \rotatebox{70}{SLD (5)}
& \rotatebox{70}{D (6)}
& \rotatebox{70}{SD (7)}
& {Mean}
& {SD}
& {Cronbach's $\alpha$} \\
\hline
PU (4)   & 15 & 25 & 0 & 0 & 0 & 0 & 0 & 1.62 & 0.49 & 0.95 \\
PEOU (4) & 23 & 15 & 2 & 0 & 0 & 0 & 0 & 1.48 & 0.60 & 0.92 \\
ATT (4)  & 11 & 21 & 8 & 0 & 0 & 0 & 0 & 1.93 & 0.64 & 0.88 \\
ITU (4)  & 13 & 20 & 7 & 0 & 0 & 0 & 0 & 1.85 & 0.69 & 0.87 \\
SN (2)   & 0 & 3 & 7 & 8 & 2 & 0 & 0 & 3.45 & 0.69 & 0.76 \\
IMG (2)  & 0 & 0 & 6 & 9 & 5 & 0 & 0 & 3.95 & 0.80 & 0.84 \\
JR (2)   & 7 & 11 & 2 & 0 & 0 & 0 & 0 & 1.75 & 0.62 & 0.89 \\
OQ (2)   & 9 & 11 & 0 & 0 & 0 & 0 & 0 & 1.55 & 0.51 & 0.91 \\
RD (2)   & 3 & 8 & 5 & 4 & 0 & 0 & 0 & 2.50 & 0.83 & 0.72 \\
\hline
\end{tabular}
\caption{TAM-based user study results showing response distributions and reliability measures (lower values = stronger agreement)}
\label{tab:tam-survey}
\end{table}

\vspace{.5em}
Perceived Usefulness ($\alpha = 0.95$, $M = 1.62$) received strong agreement, indicating participants recognized the system's value for improving model selection. Perceived Ease of Use ($\alpha = 0.92$, $M = 1.48$) showed the lowest mean, reflecting consensus that the system is intuitive and requires minimal effort.

\vspace{.5em}
Attitude Toward Using ($\alpha = 0.88$, $M = 1.93$) and Intention to Use ($\alpha = 0.87$, $M = 1.85$) both received positive evaluations, suggesting users view the system favorably and would adopt it in practice.

\vspace{.5em}
External constructs showed more diverse responses. Subjective Norm ($\alpha = 0.76$, $M = 3.45$) and Image ($\alpha = 0.84$, $M = 3.95$) yielded neutral scores, indicating that adoption is driven more by functionality than by social influence. Job Relevance ($\alpha = 0.89$, $M = 1.75$) and Output Quality ($\alpha = 0.91$, $M = 1.55$) reinforced the system's practical applicability and recommendation quality. Result Demonstrability ($\alpha = 0.72$, $M = 2.50$) showed moderate agreement with higher variance.

\vspace{.5em}
Overall, participants responded positively to HugSelect in terms of usefulness and usability. The reliability coefficients indicate consistent responses across the survey constructs, but the small sample size limits the strength of the statistical conclusions. The findings should therefore be interpreted as exploratory evidence evaluating technology acceptance and user perception.

\textbf{Summary for RQ3 (User Perception Study):} In an exploratory evaluation ($n=10$) based on TAM constructs, practitioners rated HugSelect favorably for Perceived Usefulness ($M=1.62$) and Perceived Ease of Use ($M=1.48$), confirming that structured score decomposition and explicit criterion trade-offs provide an intuitive and transparent selection aid.

\subsection{Ablation Study}
\label{subsec:ablation_study}

One advantage of the MCDM framework is its explainability, which enables analysis of how different knowledge base components influence recommendation results. During the development of HugSelect, several design variations were examined to study the effect of different feature groups and their weights. Because the framework integrates heterogeneous information sources, including metadata, functional features, and quality attributes, evaluating the contribution of each component is necessary.

\vspace{.5em}
An ablation study was conducted by selectively removing components of the HugSelect pipeline and re-evaluating system performance. This approach isolates the impact of each feature type and reveals the system's dependence on different information sources. The results in Table~\ref{tab:hugselect_ablation_significance} show that the complete HugSelect configuration achieves the best performance across both evaluation settings, confirming the benefit of combining all feature types.

\vspace{.5em}
Removing functional features causes the largest single-component drop at the model level, with Coverage@10 falling from 0.61 to 0.23 (p<0.001), indicating that functional features are the primary driver of exact-model retrieval. Removing quality attributes produces a smaller but still significant decline at the model level (0.61 to 0.41, p<0.05) and the largest relative drop at the family level (0.91 to 0.77), reflecting their complementary role in distinguishing among related model variants. When both functional features and quality attributes are removed, the system relies solely on metadata and yields the lowest performance (0.09 at ML and 0.73 at FL), underscoring the limitations of metadata-only approaches.

\vspace{.5em}
Overall, the results show that functional features and quality attributes provide complementary signals (functional features dominate exact-model retrieval, while quality attributes are especially valuable for discriminating among model families), and that combining both, together with metadata, yields the most accurate and explainable foundation-model recommendations.

\begin{table}
\centering
\begin{tabular}{lcccc}
\toprule
& \multicolumn{2}{c}{Model-Level (ML)} & \multicolumn{2}{c}{Family-Level (FL)} \\
\cmidrule(lr){2-3} \cmidrule(lr){4-5}
System Variant & Coverage@10 & p-value & Coverage@10 & p-value \\
\midrule
\textbf{HugSelect (complete)}          & \textbf{0.61} & --              & \textbf{0.91} & -- \\
w/o Functional Features   & 0.23$^{***}$  & \textless 0.001 & 0.80          & 0.13 \\
w/o Quality Attributes    & 0.41$^{*}$    & 0.01            & 0.77          & 0.07 \\
w/o Functional \& Quality & 0.09$^{***}$  & \textless 0.001 & 0.73$^{*}$    & 0.04 \\
\bottomrule
\end{tabular}
\caption{Ablation study of HugSelect at cutoff 10, evaluating the impact of removing functional features and quality attributes. Coverage@10 is reported at Model-Level (ML) and Family-Level (FL). Statistical significance is computed using the McNemar test relative to the full system; $^{*}$ $p<0.05$, $^{**}$ $p<0.01$, $^{***}$ $p<0.001$.}
\label{tab:hugselect_ablation_significance}
\end{table}

\subsection{Threats to Validity}
\label{sec:threats}

Several threats to validity should be considered when interpreting the results.

\vspace{.5em}
\textit{Construct validity.} The 44 case studies use literature-reported model choices as proxy ground truth. This is a defensible way to obtain realistic selection scenarios, but it does not imply that the reported model is the only suitable or optimal model for the task. To reduce this threat, we report both model-level and family-level metrics, because semantically appropriate alternatives may differ in version, quantization, or fine-tuning variant. Similarly, community feedback is used as evidence of perceived quality rather than as an objective measurement of runtime behavior. A review about slow inference, for example, is treated as a signal of perceived performance efficiency, not as a controlled benchmark result.

\vspace{.5em}
\textit{Internal validity.} The functional-feature and quality-attribute validation datasets were constructed through human-in-the-loop procedures supported by multiple LLMs. This design reduces single-model bias, but it does not fully remove subjectivity because feature identification and quality-attribute coding from unstructured text can be ambiguous. The ground-truth references should therefore be seen as carefully curated evaluation references rather than absolute labels. The MCDM ranking also depends on criterion weights. This is intentional because the system is designed for configurable decision support, but different weights can produce different rankings. The MoSCoW-inspired weighting scheme provides structure, yet future work should study alternative weighting strategies and compare additional MCDM methods such as AHP, TOPSIS, VIKOR, ELECTRE, PROMETHEE, and fuzzy MCDM variants.

\vspace{.5em}
\textit{Baseline validity and reproducibility.} The commercial LLM baselines were evaluated as practical zero-shot systems. This reflects common practitioner use, but it may underestimate what could be achieved with specialized prompt engineering, retrieval augmentation, repeated sampling, or fine-tuning. In addition, provider-managed systems evolve over time and do not always expose model versions, decoding settings, or retrieval behavior. For this reason, we interpret the comparison as a time-bound practical baseline rather than a fully controlled benchmark of proprietary LLMs.

\vspace{.5em}
\textit{External validity.} The knowledge base and evaluation focus on Hugging Face. Although the framework is repository-agnostic in principle, generalization to TensorFlow Hub, PyTorch Hub, ModelScope, GitHub-hosted model releases, or private enterprise model registries requires further validation. The 44 cases cover multiple domains, modalities, and tasks, but they cannot represent all software-engineering contexts in which foundation-models are selected. Domain-specific settings such as safety-critical healthcare, legal NLP, embedded deployment, and regulated industrial systems require additional evidence.

\vspace{.5em}
\textit{User-study validity.} The user study involved ten participants and should be interpreted as exploratory. The TAM-based results provide initial evidence about perceived usefulness, ease of use, output quality, and result demonstrability, but the sample size is too small for strong psychometric or population-level conclusions. Larger studies with professional developers, ML engineers, and domain stakeholders are needed to evaluate adoption, decision quality, and long-term usefulness in realistic development projects.

\vspace{.5em}
\textit{Data quality and ecosystem dynamics.} HugSelect depends on the availability and quality of model cards, README files, repository metadata, and community discussions. Sparsely documented or less popular models may receive weaker feature and quality profiles, and popular models may receive more feedback simply because they are widely discussed. The model ecosystem also changes rapidly through new releases, deprecations, fine-tuned variants, and changes in licensing. The current implementation supports periodic updates, but continuous monitoring and longitudinal validation are needed to keep the knowledge base current.

\vspace{.5em}
Despite these limitations, the evaluation provides preliminary evidence that HugSelect can support transparent foundation-model selection at repository scale. The results should be read as evidence of feasibility and practical potential, not as a final claim that the system identifies the globally optimal model for every task.

\section{Discussion}
\label{sec:discussion}

This section interprets the findings from Section~\ref{sec:evaluation} and discusses their implications for software-engineering research and practice. The central argument of this paper is that foundation-model selection should be treated as an explicit software-engineering decision rather than as an ad hoc search activity. HugSelect operationalizes this view by combining repository mining, knowledge-base construction, and multi-criteria decision support.

\subsection{Interpretation of Results}

The evaluation suggests that HugSelect can provide recommendation quality comparable to widely used conversational systems while offering a different and complementary form of support. At the family level, HugSelect achieved Coverage@10 of 0.91 and NDCG@10 of 0.74 (Table~\ref{tab:retrieval_results_extended}), indicating that the knowledge base and MCDM engine often identify model families that are consistent with literature-derived selection scenarios. At the model level, HugSelect achieved Coverage@10 of 0.61, placing it between the strongest and weaker commercial baselines in the evaluated cases.

\vspace{.5em}
The main benefit of HugSelect is not that it always outperforms conversational recommenders. Rather, the results show that explicit decision logic can reach competitive retrieval performance while providing traceability that general-purpose LLM interfaces do not provide by default. The score decomposition, feature matching, quality-attribute links, and family-level mappings make it possible to inspect why a model is recommended and which trade-offs drive the ranking. This is particularly important when model selection must be justified to project teams, clients, reviewers, auditors, or other stakeholders.

\vspace{.5em}
The ablation study reinforces this interpretation. Removing functional features produced the largest drop in model-level Coverage@10, while removing quality attributes also reduced performance and weakened the ability to discriminate among related variants. This indicates that model selection benefits from combining multiple evidence sources rather than relying only on metadata or popularity. At the same time, the modest Precision@10 scores across all systems show that foundation-model selection remains exploratory: a useful system should help users inspect a set of plausible candidates, not merely output a single definitive answer.

\vspace{.5em}
The user study provides preliminary evidence that practitioners value this structured form of decision support. Participants rated perceived usefulness, ease of use, job relevance, and output quality positively (Table~\ref{tab:tam-survey}). Because the study is small, these findings should be interpreted as early usability evidence rather than conclusive proof of adoption. Still, the results support the claim that explicit criteria and explanations can make foundation-model selection more understandable.

\subsection{Implications for Software Engineering Practice}

\paragraph{Model selection as component selection.}
foundation-models increasingly function as reusable AI components in software systems. Selecting such a component affects architecture, integration effort, deployment cost, license compliance, quality risks, and maintainability. HugSelect makes these concerns explicit by representing models as alternatives with metadata, functional capabilities, and perceived quality attributes.

\paragraph{Traceable recommendations.}
In practical software projects, a recommendation is useful only if stakeholders can understand and challenge it. The MCDM-based ranking mechanism supports this need by exposing criterion-level score contributions and the evidence behind them. This makes the recommendation process more auditable than a purely conversational answer, while still allowing users to adjust priorities.

\paragraph{Repository-scale knowledge integration.}
The results show that repository metadata alone is insufficient for effective selection. Model cards, README files, and community feedback contain complementary information that can improve matching and explanation. HugSelect therefore demonstrates how repository mining can be connected to decision-support models for AI-enabled system development.

\subsection{Implications for Research}

For software-engineering research, HugSelect contributes evidence that MCDM-based technology-selection frameworks can be extended from traditional software and cloud-service selection to foundation-model ecosystems. The work also highlights new research challenges: maintaining current knowledge bases for rapidly changing model hubs, validating perceived quality signals against runtime benchmarks, evaluating recommendations when multiple alternatives are acceptable, and designing interactive tools that help users negotiate trade-offs.

\subsection{Future Directions}

\paragraph{Richer evidence sources.}
Future versions should incorporate benchmark results, fine-tuning datasets, hardware requirements, energy consumption, license-change histories, vulnerability reports, and deployment documentation. These sources would help connect perceived quality signals to more objective evidence.

\paragraph{Interactive refinement.}
HugSelect can be extended with conversational refinement, relevance feedback, and what-if analysis so users can iteratively adjust criteria and inspect how rankings change under different priorities.

\paragraph{Cross-repository integration.}
Although this study focuses on Hugging Face, the architecture can be adapted to other repositories and private model registries. Cross-repository integration would support organizations that combine public open-source models with internal fine-tuned variants.

\paragraph{Longitudinal and industrial evaluation.}
Future work should evaluate HugSelect in longitudinal industrial settings, where teams use the system during real development projects and later assess whether the selected models satisfy operational requirements over time.

\section{Conclusion}
\label{sec:conclusion}

This paper addressed the problem of selecting foundation-models from large and heterogeneous repositories. We argued that this problem should be treated as a software-engineering decision because foundation-models increasingly act as reusable AI components whose selection affects system functionality, quality attributes, deployment constraints, licensing, cost, and maintainability.

HugSelect was proposed as an explainable multi-criteria decision-support framework for this setting. The framework integrates structured repository metadata, functional features extracted from model-card descriptions, and perceived quality signals derived from community feedback into a unified knowledge base. It then applies a weighted WSM/SAW decision model to rank candidate models and expose criterion-level contributions to the final score.

The evaluation provides preliminary but encouraging evidence. The extraction-pipeline validation shows that useful functional and perceived quality signals can be obtained from heterogeneous repository and community sources, although imperfectly. The 44-case comparative evaluation shows that HugSelect achieves recommendation quality competitive with practical commercial LLM-based baselines, especially at the family level, while providing more explicit and auditable reasoning. The ablation study demonstrates that functional features and perceived quality attributes contribute complementary value beyond metadata alone. The user study suggests that participants perceive the system as useful, understandable, and relevant, although the small sample size limits the strength of this conclusion.

The main contributions of this paper are: (i) a repository-scale MCDM formulation of foundation-model selection for AI-enabled software engineering; (ii) a knowledge-base construction pipeline integrating metadata, model descriptions, and community-derived perceived quality evidence; (iii) an explainable ranking engine with criterion-level score decomposition; and (iv) an empirical evaluation combining extraction validation, comparative case studies, ablation analysis, and an exploratory user study.

Several limitations remain. Literature-reported model choices provide realistic but imperfect proxy ground truth, community feedback reflects perceived rather than objective quality, commercial LLM baselines are time-bound, and the user study is exploratory. Future work should expand the evaluation to additional repositories and industrial contexts, incorporate objective benchmark and deployment evidence, support interactive refinement, and study how recommendations evolve as model ecosystems change.

Overall, HugSelect demonstrates that repository mining and multi-criteria decision making can be combined to support more transparent foundation-model selection. By exposing evidence and trade-offs rather than hiding them inside a single opaque recommendation, the framework offers a practical basis for informed model selection in software engineering and AI-enabled system development.

\section*{Declaration of Competing Interest and Data Availability}
The authors declare that they have no known competing financial interests or personal relationships that could have appeared to influence the work reported in this paper. 

\vspace{1em}
\noindent \textbf{Data Availability} \\
Data associated with this study have been deposited at Mendeley Data under the accession number 10.17632/9kxtkvyv5m.2 (\url{https://doi.org/10.17632/9kxtkvyv5m.2}). The dataset includes the raw model information, validation data, case study results, user study records, and data collection and processing code.

\appendix

\section{Literature Study Gap-Analysis Table}
\label{app:gap-analysis}

Table~\ref{tab:related-work-comparison} presents the full gap-analysis matrix used to position HugSelect relative to related work. The original visual matrix has been converted into grouped descriptors to improve readability.

\begin{landscape}
\scriptsize
\setlength{\tabcolsep}{2pt}
\renewcommand{\arraystretch}{1.18}
\begin{longtable}{L{2cm}cL{2.0cm}ccL{1.75cm}L{2.55cm}L{1.45cm}L{2.5cm}L{1.45cm}L{1.85cm}L{1.75cm}L{1.5cm}}
\caption{Comparison of studies on decision-making and recommendation support for foundation-model-related domains.}\label{tab:related-work-comparison}\\
\toprule
\textbf{Study} & \textbf{Year} & \textbf{Domain} & \textbf{Alt.} & \textbf{Crit.} & \textbf{Approach} & \textbf{Data sources} & \textbf{DC mode} & \textbf{Feature / evidence collection} & \textbf{Decision model} & \textbf{MCDM method} & \textbf{Evaluation} & \textbf{Weighting}\\
\midrule
\endfirsthead
\toprule
\textbf{Study} & \textbf{Year} & \textbf{Domain} & \textbf{Alt.} & \textbf{Crit.} & \textbf{Approach} & \textbf{Data sources} & \textbf{DC mode} & \textbf{Feature / evidence collection} & \textbf{Decision model} & \textbf{MCDM method} & \textbf{Evaluation} & \textbf{Weighting}\\
\midrule
\endhead
\midrule
\multicolumn{13}{r}{\emph{Continued on next page}}\\
\endfoot
\bottomrule
\multicolumn{13}{p{24.0cm}}{\textit{Notes.} Alt. = number of alternatives; Crit. = number of criteria; DC mode = data collection mode; KB/ED = knowledge-based or expert-driven decision model; DDDM = data-driven decision-making; WSM/SAW = Weighted Sum Model / Simple Additive Weighting. The original visual matrix used x-marks; this table converts the same information into grouped descriptors to improve readability.}\\
\endlastfoot
Leščinskaitė, P. et al. & 2026 & ML Artifacts & 139 & 7 & Overview, Benchmarking & Public literature, OpenAI & Semi-automated & Textual attributes, Benchmarking & MCDM & WSM/SAW, AHP, TOPSIS & Comparative, Other & --\\
Olabanjo, O. et al. & 2026 & AI System Designs & 10 & 8 & Overview, Selection & Public literature & Automated & Surveying, Expert opinion, Benchmarking & MCDM & WSM/SAW, AHP, TOPSIS, Other & Comparative, Other & Value-based, Statistical\\
Farshidi, S. et al. & 2025 & Software Packages & 39,841 & 3 & Overview, Taxonomy, Selection & GitHub, Public literature, Public dataset, Stack Overflow, Gray literature & Automated & Metadata extraction, Textual attributes, Entity/relation extraction, Surveying, Standard mapping, Machine learning & MCDM & WSM/SAW, Fuzzy MCDM & Comparative, User study, Qualitative & Graph-based, Value-based\\
Chen, Q. et al. & 2025 & foundation-models & 121,404 & 14 & Taxonomy, Selection, Benchmarking & Hugging Face & Automated & Metadata extraction, Textual attributes, Entity/relation extraction & KB/ED, DDDM & -- & Comparative & AI-based\\
Mienye, I. et al. & 2025 & Large Language Models & 31 & 4 & Overview, Taxonomy & Public literature & -- & Surveying & None & -- & None & --\\
Minaee, S. et al. & 2025 & Large Language Models & 48 & 3 & Overview, Taxonomy, Benchmarking & Public literature, Public dataset & -- & Benchmarking & None & -- & None & --\\
Radulescu, C. et al. & 2025 & GenAI Models & 5 & 7 & Selection & Public literature & -- & -- & MCDM & Other & Other & --\\
Gonzalez, A. et al. & 2025 & Pre-Trained Models & 2,205 & 5 & Taxonomy, Selection & Public literature, Hugging Face & -- & Metadata extraction, Textual attributes & KB/ED & -- & None & --\\
Jain, A. et al. & 2025 & Pre-Trained Models & 5 & 4 & Selection, Benchmarking & Public literature & Manual & Metadata extraction, Textual attributes, Machine learning, Benchmarking & MCDM & TOPSIS & Comparative, Other & --\\
Alsalem, M.A. et al. & 2024 & AI Applications & 50 & 7 & Selection, Benchmarking & Public literature & Manual & Expert opinion, Standard mapping & MCDM & Fuzzy MCDM, Other & Other & Value-based\\
Drissi, N. et al. & 2024 & ML Models & 8 & 7 & Overview, Benchmarking & Public literature & Manual & Textual attributes, Expert opinion, Benchmarking & AI/ML, MCDM & WSM/SAW, AHP & Comparative, Other & Value-based\\
Bhol, S.G. et al. & 2024 & MLaaS & 4 & 17 & Selection & Public literature, Gray literature & Manual & Metadata extraction, Textual attributes, Entity/relation extraction, Surveying & MCDM & AHP, TOPSIS & Comparative & Value-based\\
Di Sipio, C. et al. & 2024 & Pre-Trained Models & 3 & 6 & Taxonomy, Selection & Public literature, Hugging Face & Semi-automated & Metadata extraction, Textual attributes & KB/ED & -- & User study & Value-based\\
Suryani, M. et al. & 2024 & foundation-models & 1,090 & 2 & Overview, Taxonomy & GitHub, Public literature, Hugging Face & -- & Metadata extraction, Entity/relation extraction & None & -- & None & --\\
Chen, Z. et al. & 2024 & foundation-models & 43 & 6 & Overview & Public literature & -- & Surveying & None & -- & None & --\\
Zhou, J. et al. & 2024 & Agents & 97 & 9 & Taxonomy & Public literature, Gray literature & Manual & Surveying & None & Other & Qualitative & --\\
Ding, Y. et al. & 2024 & Vision-Language Models & 17 & 10 & Selection, Benchmarking & Public dataset & Manual & Machine learning, Benchmarking & AI/ML, DDDM & Other & Comparative & --\\
Lu, Q. et al. & 2023 & foundation-models & 72 & 3 & Taxonomy & Public literature & -- & Surveying & None & -- & None & --\\
Azad, B. et al. & 2023 & foundation-models & 40 & 22 & Overview, Taxonomy & Public literature & Manual & Surveying & None & -- & None & --\\
Awais, M. et al. & 2023 & foundation-models & 97 & 9 & Overview, Taxonomy & Public literature & Manual & Surveying & None & -- & None & --\\
Shen, Y. et al. & 2023 & foundation-models & 23 & 3 & Selection & GitHub, Hugging Face, OpenAI & -- & -- & AI/ML & -- & Comparative & --\\
Zhou, C. et al. & 2023 & foundation-models & 67 & 3 & Overview, Taxonomy & Public literature & -- & Surveying & None & -- & None & --\\
Chakrabortty, R. et al. & 2023 & Chatbots & 8 & 9 & Selection & Public literature & Manual & Expert opinion & MCDM & TOPSIS, Fuzzy MCDM & Other & Value-based\\
Hu, Z. et al. & 2023 & Pre-Trained Models & 6 & 3 & Selection, Benchmarking & GitHub & Manual & Benchmarking & AI/ML & Other & Comparative & --\\
Zhao, W. et al. & 2023 & Large Language Models & 57 & 5 & Overview & Public literature, Stack Overflow & -- & Surveying & None & -- & None & --\\
Tsay, J. et al. & 2023 & Machine Learning Models & 7,900 & 8 & Overview & GitHub, Public literature, Public dataset & Automated & Metadata extraction & DDDM & -- & Comparative & --\\
Liu, M. et al. & 2023 & Machine Learning Models & 25,718 & 11 & Taxonomy, Selection & GitHub, Stack Overflow & -- & Textual attributes, Entity/relation extraction & KB/ED & -- & User study & --\\
Khan, S. et al. & 2022 & Transformer Models & 60 & 6 & Overview & Public literature & -- & Surveying & None & -- & None & --\\
Şeker and Kahraman & 2021 & Software Packages & 4 & 8 & Selection & Public literature & Manual & Expert opinion & MCDM & TOPSIS, Fuzzy MCDM & Comparative, Other & Performance-based, Statistical\\
Cao, X. et al. & 2021 & Machine Learning & 6,239 & 3 & Selection, Benchmarking & GitHub, Public dataset & -- & Textual attributes, Entity/relation extraction & AI/ML, DDDM & -- & Comparative & Graph-based\\
\rowcolor{gray!12}
\textbf{HugSelect} & \textbf{2026} & \textbf{foundation-models} & \textbf{71,274} & \textbf{3} & Overview, Taxonomy, Selection & GitHub, Public literature, Public dataset, Hugging Face, Stack Overflow, Gray literature & Automated & Metadata extraction, Textual attributes, Entity/relation extraction, Surveying, Expert opinion, Standard mapping, Machine learning & MCDM & WSM/SAW & Comparative, User study, Qualitative & Graph-based, Value-based\\
\end{longtable}
\end{landscape}

\bibliographystyle{cas-model2-names}

\bibliography{base}

\end{document}